\documentclass[aps,prb,twocolumn,superscriptaddressgroupaddress,floatfix]{revtex4-2}
\usepackage{amssymb}
\usepackage{graphicx}
\usepackage{times}
\usepackage{amsmath}
\usepackage{amsthm}
\usepackage{amsfonts}
\usepackage[T1]{fontenc}
\usepackage[latin9]{inputenc}
\usepackage{array}
\usepackage{multirow}
\usepackage{color}
\usepackage{bm}
\usepackage{color}

\usepackage{bbm}
\usepackage{hyperref}
\usepackage{babel}
\usepackage[mathscr]{euscript}
\usepackage{float}
\usepackage{hyperref}
\hypersetup
{	colorlinks,%
	citecolor=green,%
	linkcolor=red,%
	urlcolor=blue%
}
\allowdisplaybreaks[4]

\begin{document}
\title{Role of many phonon modes on the high-temperature linear-in-\texorpdfstring{$T$}{T} electronic resistivity}
\author{Sankar Das Sarma}
\author{Yi-Ting Tu}
\affiliation{Condensed Matter Theory Center and Joint Quantum Institute, Department of Physics, University of Maryland, College Park, Maryland 20742, USA}

\begin{abstract}
  We theoretically consider the possibility that phonons may be playing a role in the observed linear-in-$T$ resistivity in cuprates by focusing on the obvious question: How can phonon scattering be consistent with a linear-in-$T$ resistivity with a constant slope given that cuprates have many phonon modes with different energies and electron-phonon couplings (e.g.\ 21 phonon modes for LSCO)?  We show using an arbitrarily large number of independent phonon modes that, within a model Boltzmann transport theory, the emergent high-$T$ linear-in-$T$ resistivity manifests an approximately constant slope independent of the number of phonon modes except in some fine-tuned narrow temperature regimes.  We also comment on the quantitative magnitude of the linear-in-$T$ resistivity in cuprates pointing out the constraints on the effective electron-phonon coupling necessary to produce the observed resistivity.
\end{abstract}

\maketitle

\section{Introduction}

A central topic in the physics of cuprates over the last 30 years has been its so-called ``strange metallicity''. An important component of this ``strangeness'' is a linear-in-$T$ resistivity over a large temperature range, particularly in the optimally doped regime~\cite{Greene2020,Hartnoll2022,Phillips2022,Zaanen2019,Legros2019}.  The effect is most pronunced near optimal doping ($x\sim 0.15$) in $\text{La}_{2-x}\text{Sr}_x\text{CuO}_4$ (LSCO), where the measured resistivity, $\rho(T)$, is linear from $\sim 50\,\text{K}$ to the highest measurement temperature $\sim 1000\,\text{K}$, with a constant slope $d\rho/dT$ throughout~\cite{Takagi1992,Komiya2005}.  Although such an impressive linear-in-$T$ resistivity seems to be specific to the fine-tuned doping $x\sim 0.15$ in LSCO, similar, but not as spectacular, behavior of a sustained approximate linear-in-$T$ resistivity appears generic in many (but, by no means all) cuprates (both with hole and electron doping).

The current work is motivated by the specific question of whether such a linear-in-$T$ resistivity could arise at all, even as a matter of principle, from phonon scattering effects~\cite{Wu2019,DasSarma2020,Hwang2019,DasSarma2022,Sarkar2018}, which is generally discarded by the experts as a possible mechanism for strange metallicity in cuprates~\cite{Greene2020,Hartnoll2022,Phillips2022,Zaanen2019,Legros2019}. We emphasize that ``strange metallicity'' is not a sharp concept and consists often of a collection of phenomena, of which a linear-in-$T$ resistivity is perhaps the most discussed aspect.  We focus \emph{only} on the linear-in-$T$ resistivity and are not discussing any other aspects of ``strange metallicity''~\cite{Greene2020,Hartnoll2022,Phillips2022,Zaanen2019,Legros2019}. The goal here is not by any means to claim that phonon scattering is necessarily leading to the observed ``strange metallicity'' in cuprates, but to critically consider whether there are compelling reasons to assert that phonons cannot possibly be playing any role in this physics of linear-in-$T$ resistivity at all.  This is not an unreasonable goal because (1) no commonly accepted mechanism for this linear-in-$T$ behavior in cuprates is agreed upon by the community in spite of hundreds of papers focused on this topic, and (2) most normal metals (e.g.\ Al, Ag, Cu, Pb) also manifest a linear-in-$T$ resistivity over a large temperature range ($50\,\text{K}$--$800\,\text{K}$) arising entirely from phonon scattering in the high-$T$ equipartition regime ($>T_D/5$, where $T_D$ is the Debye temperature)~\cite{Ziman2001,Ashcroft1976,Grimvall1976,Hwang2019,Min2012}. We mention that our focus is entirely on the linear-in-$T$ resistivity behavior and no other aspects of strange metallicity. To avoid any misunderstanding, we emphasize that our focus is entirely on the zero magnetic field situation, which we believe needs to be explained at some level before the more complex situation of magnetotransport in the presence of a magnetic field could be considered.  The linear-in-$T$ behavior we focus on thus does not persist to arbitrarily low temperatures, applying generally for $T>T_c$ in cuprates (where $T_c$ is the superconducting transition temperature) and for $T>50\,\text{K}$ in most metals.

\section{Resistivity due to phonons}

The linear-in-$T$ resistivity in metals arising from phonons with a dimensionless electron-phonon coupling of $\lambda$ and a Debye temperature (or typical phonon energy) $T_{D}$ follows from the well-established Bloch-Gr\"uneisen transport theory for carrier transport limited by phonon scattering~\cite{Ziman1972,Allen1999,Kawamura1992,Allen1978}.  In particular, the standard transport theory gives the following leading order metallic resistivity $\rho(T)$ at high temperatures ($T\gg T_D$) arising from electron-phonon scattering in 3D systems, where $\tau$ is the electron-phonon transport scattering time~\cite{Allen1988}
\begin{equation}\label{eq:rhoT}
  \rho(T)=\frac{4\pi}{\omega_{p}^{2}\tau(T)},
\end{equation}
\begin{equation}\label{eq:tauT}
  \frac{\hbar}{\tau(T)}=2\pi k_{B} T \lambda \left[1-\frac{T_D^2}{12T^{2}}\right].
\end{equation}
Here $\omega_p={(4\pi n e^2/m)}^{1/2}$ is the effective plasma frequency of the metal, defined by the effective carrier density $n$ and the effective mass $m$, and $T_D$ is the Debye temperature (or the typical phonon temperature scale). Here, Eq.~(\ref{eq:tauT}) arises from an expansion in $T_D/T$ (valid for ``high temperatures'') where subleading terms in higher orders, $O(T_D/T)^4$ and above, have been neglected. Note that the factor of $12$ in Eq.~(\ref{eq:tauT}) may depend on the details such as the exact form of the dispersions of electrons and phonons, but the leading high-temperature ($T\gg T_D$) correction to linear-in-$T$ goes as $O(T_D/T)^2$ quite generally~\cite{Allen1988}.  In our work, we do not use such a high-temperature expansion anywhere, and we provide Eqs.~(\ref{eq:rhoT})--(\ref{eq:tauT}) only for the sake of completeness. An important feature of this $T_D/T$ expansion is that the linear-in-$T$ resistivity is an asymptotic high-$T$ result, and there is really no sharp $T$-value where suddenly the resistivity becomes linear-in-$T$.  The corresponding low-$T$ resistivity (for $T\ll T_D$) is also well-known, going as $T^5$ or $T^4$ depending on the system dimensionality being 3 or 2 respectively.  We mention that these approximations apply when the phonon scattering is quasi-elastic, which is a well-valid approximation when the phonon frequency is much smaller than the electronic Fermi energy, which is the situation we consider. From this high-$T$ expansion, we see that for ${(T_D/T)}^2<12$, the metal manifests an approximate linear-in-$T$ resistivity---explicit calculations show that the linear-in-$T$ resistivity applies for $T>T_D/5$, continuing indefinitely for higher $T$, thus explaining the linear-in-$T$ high-$T$ resistivity in normal metals. We mention that the bound $T_D/5$ should not be taken too seriously as the precise bound varies between $\sim T_D/3$--$T_D/7$ depending on the other parameters of the problem, but the main physics is that the linearity sets in at a temperature much below $T_D$.  In addition, as discussed later, there is no sharp onset temperature for the linearity to show up since the resistivity crosses over from a power law with high power at low $T$ to a linear-in-$T$ resistivity over a temperature range as should be obvious from Eq.~(\ref{eq:tauT}) above.  We note that the results defined by Eqs.~(\ref{eq:rhoT}) and (\ref{eq:tauT}) are standard and can be found in the literature going back 50--80 years, and thus a linear-in-$T$ resistivity is the generic result for phonon induced metallic resistivity for $T>T_D$~\cite{Ziman1972,Allen1999,Kawamura1992}. For quantitative comparison with experiments, of course, sophisticated (and completely numerical) first-principles calculations are necessary using the detailed appropriate band structures and solving the Boltzmann equation numerically~\cite{Macheda2018,Ponce2021}.

We note that for materials with low Fermi momentum $k_F$ (or equivalently, low carrier density $n$, since $k_F \sim n^{1/d}$, with $d$ being the system dimensionality), $T_D$ ($>T_{BG}$) is replaced by $T_{BG}=2\hbar k_F v$, where $v$ is the sound (or phonon) velocity~\cite{Hwang2019}.
Note that $T_{BG} <2T_D$ in general (since the maximal possible $k_F$ is the size of the Brillouin zone and $T_D$ corresponds to the maximum phonon energy with a phonon wave vector equal to the Brillouin zone boundary), but in principle $T_{BG}$ could be much less than $T_D$ in low-density metals, leading to the linear-in-$T$ behavior persisting to low temperatures, depending on how low the effective carrier density is, defining $T_{BG} \sim n^{1/d}$~\cite{Wu2019,DasSarma2020,Hwang2019,DasSarma2022,Sarkar2018,Davis2023}.  We stick to the situation (as in normal metals) where $T_{BG}>T_D$, and do not consider low-density metals although such a situation may be relevant to under-doped cuprates with small Fermi surfaces.

The key thing to focus on for our purpose here is that, according to Eq.~(\ref{eq:tauT}) above, the phonon-induced resistivity is linear-in-$T$ for $T>T^*\sim T_D/5$ with the slope $d\rho/dT \sim \lambda$.  In a system with many phonon modes (e.g.\ LSCO has 21 phonon modes and YBCO has 38), there are in principle as many $\lambda$ and $T_D$ values as there are phonon modes.  Therefore, one could argue that a linear-in-$T$ resistivity with an apparent constant slope is unlikely in a system with many phonon modes since different phonon modes would manifest different $T^*$ with different $\lambda$-dependent slopes resulting in a non-linear (in $T$) resistivity even at high temperatures.  Thus, the observed linear-in-$T$ resistivity in cuprates may seem qualitatively inconsistent with phonon scattering, given the large number of phonon modes in a material with multi-atom unit cells.  This qualitative issue, the main topic of the current work, is independent of any quantitative consideration of the actual magnitude of the resistivity (or the slope $d\rho/dT$) which is proportional to the electron-phonon coupling (and other electronic band parameters such as the effective plasma frequency).
The basic question addressed here is whether the resistivity limited by scattering from many phonon modes in a system leads to nonlinear (in $T$) behavior at high $T$ instead of the linear-in-$T$ phonon-induced metallic resistivity generic in normal metals.

The quantity to calculate for the resistivity, according to Eq.~(\ref{eq:rhoT}), is the transport relaxation time $\tau$, which for a single phonon mode is easily calculated within the leading-order Boltzmann theory, leading to the following well-known Bloch-Gr\"uneisen resistivity formula arising from electron-phonon interaction:
\begin{multline}\label{eq:rhoBG}
  \rho(T) = \rho_0 +\\
  \frac{2\pi\lambda k_B T/\hbar}{(n/m)e^2}\int_0^{\omega_D} \frac{d\omega}{\omega}{\left(\frac{\omega}{\omega_D}\right)}^4{\left[\frac{\hbar\omega/k_B T}{\sinh(\hbar\omega/2k_B T)}\right]}^2.
\end{multline}

In Eq.~(\ref{eq:rhoBG}), $n$ and $m$ are effective carrier density and effective mass with $\omega_D$ being the Debye frequency (or temperature, with $T_D =\hbar \omega_D/k_B$) imposing a high-energy cutoff on the phonon spectrum (For systems with small Fermi surfaces, where $T_D>T_{BG}$, the cutoff $\omega_D$ is replaced by $\omega_{BG}=k_B T_{BG}/\hbar$ with no loss of generality, affecting only the quantitative issue of how low in $T$ the linearity should persist in general).  The first term, $\rho_0$, is the sample-dependent nonuniversal temperature-independent residual resistivity, arising from elastic disorder/defect/impurity scattering, which is of no consequence in the current work and is ignored in the rest of this work and in our results, where we set $\rho_0=0$ with no loss of generality.  When the phonons are described by the Einstein model with one sharp frequency, $\omega_E$, instead of a phonon dispersion as in the Debye model, the $(\omega/\omega_D)^4$ factor inside the integral gets replaced by $(\omega_E/4) \delta (\omega-\omega_E)$, making the frequency integral trivial.  We note that the thermal factor within the square bracket in Eq.~(\ref{eq:rhoBG}) implies the asymptotic temperature dependence of the phonon-induced resistivity as being $T$ (or $T^5$) for high (or low) temperatures with the temperature scale defining high versus low $T$ behavior, according to Eq.~(\ref{eq:tauT}), being approximately $\sim T_D/5$ where $T_D$ is the Debye temperature ($T_D = \hbar\omega_D/k_B$).

\begin{figure*}
  \includegraphics[trim=12 0 0 0, scale=0.94]{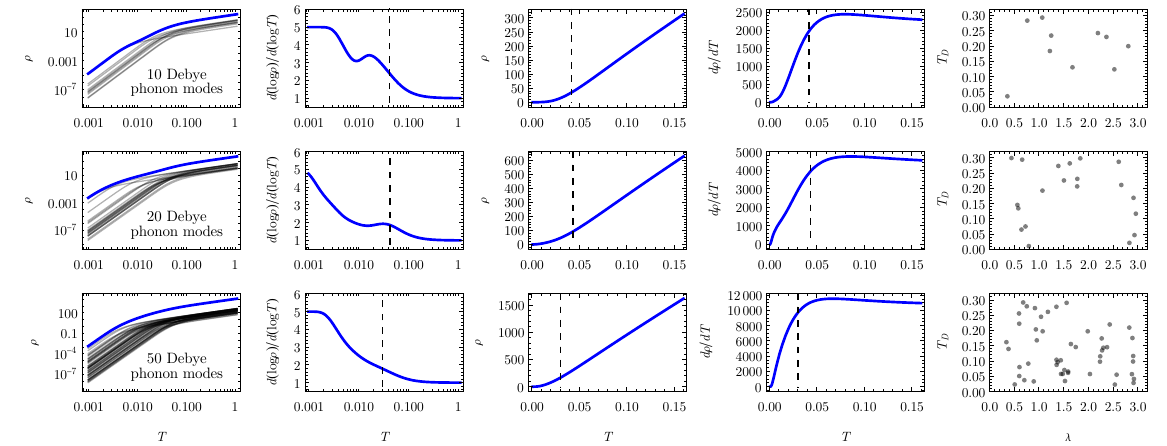}
  \caption{Calculated resistivity as a function of $T$ due to 10 (top row), 20 (middle row), and 50 (bottom row) Debye phonon modes with randomly chosen parameters shown in the right column. Gray curves indicate the resistivity contribution from individual phonon modes, and blue indicates the total resistivity. Vertical dashed lines indicate $T^{*}$, the temperature at which $d\rho/dT$ becomes approximately constant. The units are arbitrary. }
  \label{fig:Debye}
\end{figure*}

We note that, although we take the electron-phonon coupling constant $\lambda$ as a parameter, in principle it is determined by the electronic structure and phonon details of the material through the formula~\cite{Allen1999}
\begin{equation}\label{eq:lambda}
  \lambda = N_F \frac{\sum_{k,k'}(v_{kx}-v_{k'x})^2 |M_{k,k'}|^2/\hbar\omega_{k-k'}\delta(\epsilon_k)\delta(\epsilon_{k'})}{\sum_{k,k'}(v_{kx}-v_{k'x})^2\delta(\epsilon_k)\delta(\epsilon_{k'})}.
\end{equation}
Here, $N_F$ is the electronic density of states on the Fermi surface, $M$ is the electron-phonon scattering matrix elements between electron wavenumbers $k$ and $k'$, $\omega_{k-k'}$ is the relevant phonon frequency for the momentum exchange $k-k'$, $v_{kx}$ is the electron group velocity  $\partial\epsilon_k/\partial k_x$, where $\epsilon_k$ is the band energy dispersion.  We note that the above equation for $\lambda$ applies to each phonon mode, and the total or net effective electron-phonon coupling, defining the transport scattering rate and carrier resistivity, would be a combined effect of all the electron-phonon couplings in the material, which could considerably enhance its effective strength.

\section{Effect of many phonon modes}

To ascertain the effect of many phonon modes in the system, we must take into account the sum of the resistivity arising from each individual phonon mode.  In doing so, the Boltzmann transport theory asserts that one must carry out the thermal average for the scattering time before calculating the resistivity, which is proportional to $1/\tau$ (and not to $\tau$).  This may lead to a violation of Matthiessen's rule for temperatures comparable to the Fermi temperature.  Since we restrict ourselves to $T\ll T_F$ for all our results, we find that Matthiessen's rule applies well for all the results presented in this work, since the correction to Matthiessen's rule is suppressed by $\sim O(T/T_F)^2$.  (We explicitly checked that Mattheissen's rule is well-valid for all the results shown in this paper.)  Thus, by calculating the resistivity $\rho_n(T)$ arising from the individual phonon mode, defined by individual phonon frequency $\omega_{D,n}$ (or $\omega_{E,n}$) and individual electron-phonon coupling $\lambda_n$, the total resistivity is simply calculated by the sum $\rho(T)=\sum_n \rho_n(T)$ over the individual contributions.

\begin{figure*}
  \includegraphics[trim=12 0 0 0, scale=0.94]{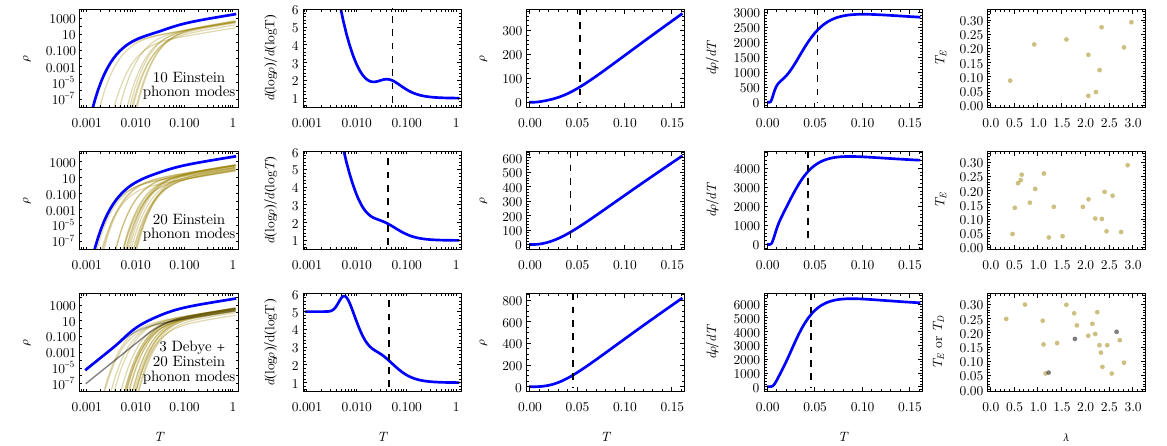}
  \caption{Calculated resistivity as a function of $T$ due to 10 (top row) and 20 (middle and bottom row) Einstein phonon modes, with the addition of 3 Debye phonon modes in the bottom row, all with randomly chosen parameters. Gold (gray) curves indicate the resistivity contribution from individual Einstein (Debye) phonon modes, with the corresponding parameters shown in the right panel in the respective colors, and blue indicates the total resistivity. Vertical dashed lines indicate $T^{*}$, the temperature at which $d\rho/dT$ becomes approximately constant. The units are arbitrary. }
  \label{fig:Einstein}
\end{figure*}

Since we keep all other system variables (e.g.\ $n$, $m$, etc.) fixed, the final resistivity depends only on the phonon variables, i.e., on the set of individual phonon frequencies and coupling constants.  We take the Fermi temperature $T_F$ to be much larger than the characteristic phonon frequencies (e.g.\ the Debye or the Einstein temperature) throughout with no loss of generality since our interest is on the $T$-dependence arising from the phonon physics, in particular, the role of multiple distinct phonon modes on the high-$T$ (but still $<T_F$) linear-in-$T$ electronic resistivity. Note that rescaling $T$ and $T_D$ in Eq.~(\ref{eq:rhoBG}) only leads to a rescaling of $\rho$, so the result will only depend on $T/T_D$ and the unit of $T$ can thus be chosen arbitrarily. In the calculation below, we use arbitrary units for temperature and resistivity that correspond to setting $m$, $e$, $T_{F}$, $k_{B}$, $\hbar$ in Eq.~(\ref{eq:rhoBG}) to the unity.

The theory now involves using a large number of individual phonon modes with arbitrary $\omega_D$ and $\lambda$, and calculating the resistivity.  Since the resistivity slope in temperature, $d\rho_n/dT$, in the ``high-temperature'' linear-in-$T$ equipartition regime arising from the $n$th phonon mode is given by $\lambda_n$, and since the linearity sets in at a temperature scale $T \sim T_{D,n}/5$, where $\lambda_n$ and $T_{D,n}$ are the coupling and Debye temperature for the $n$th mode, we expect the net resistivity summing over all contributing $\rho_n$ to look nonlinear in the whole temperature range.  This naive expectation is, however, not what happens when the resistivity calculation is carried out in actuality as described below, particularly when the number of phonon modes is large ($\gg1$), which is the case for cuprates.

In Fig.~\ref{fig:Debye}, we show the calculated resistivity as a function of temperature (with fixed $T_F$, whose exact value is not important as long as it is much larger than the temperature range being considered) for scattering from 10, 20, and 50 completely random Debye phonon modes, with each phonon mode having random coupling and random $T_D$.  The actual random values of the individual $\lambda$ and $T_D$ are also shown in each figure.  No significance should be attached to the choice of parameters and the magnitude of the resistivity---the only goal here is to investigate the linearity or not of the effective ``high-$T$'' resistivity in the presence of a large number of phonon modes with arbitrary coupling and dispersion.  In each case, the individual resistivity arising from each mode is shown (in gray) along with the net resistivity from all the modes together (in blue, which would be the experimental situation since the carriers only care about the total resistive scattering from all mechanisms).  The contribution of each mode shows the usual ``bending'' behavior from a high power law ($\sim T^5$) at low-$T$ ($T<T_D/5$) to the linear-in-$T$ high-$T$ ($T>T_D/5$) behavior with the resistivity slope at high-$T$ being proportional to the corresponding $\lambda$.  The net resistivity, however, shows some low-$T$ crossover behavior to a high-$T$ linear behavior at some $T^{*}$, and the linearity, once it sets in for, continues to arbitrary $T>T^{*}$ within our Boltzmann-Debye transport theory.  This persistence of a clearly linear high-$T$ resistivity happens for all three cases with 10, 20, 50 phonon modes in Fig.~\ref{fig:Debye}.  We have studied many examples with many more phonon modes, always finding a well-defined linear-in-$T$ behavior arising from electron-phonon interactions independent of how many phonons with random coupling and dispersion are used in the calculation.

\begin{figure*}
  \includegraphics[trim=12 0 0 0, scale=0.94]{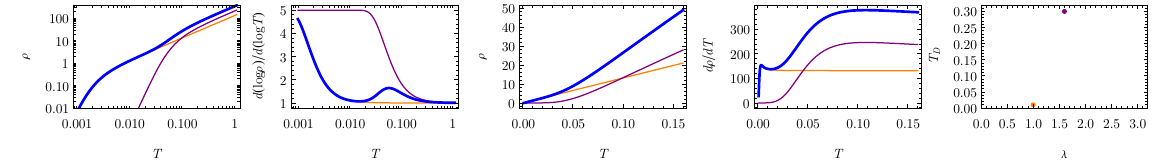}
  \caption{Calculated resistivity as a function of $T$ due to two Debye phonon modes with parameters chosen such that $\rho(T)$ shows two linear segments (nearly constant $d\rho/dT$) with different slopes. Orange and purple curves indicate the contribution from the two individual phonon modes, with the parameters shown in the right panel in the respective colors, and blue indicates the total resistivity. The units are arbitrary. }
  \label{fig:artificial}
\end{figure*}

In the top and middle row of Fig.~\ref{fig:Einstein}, the calculated resistivity for phonon scattering in the Einstein model, again with random electron-phonon coupling and phonon frequency, are shown for 10 and 20 random Einstein modes.  Here, the low-$T$ resistivity is exponential in temperature, but the high-$T$ equipartition behavior is still linear-in-$T$ since the phonons become classical in the high-$T$ regime.  We note that the qualitative high-$T$ behavior for Debye (Fig.~\ref{fig:Debye}) and Einstein (Fig.~\ref{fig:Einstein}) models are the same, and above a transient crossover low-$T$ regime, both models show manifest linear-in-$T$ resistivity independent of how many phonon modes are contributing to the resistivity with an effective slope determined by the random phonon couplings.

In the bottom row of Fig.~\ref{fig:Einstein}, we show results using both models together, using 20 random Einstein modes and 3 random Debye modes.  (The results are representative as we have produced results combining many more situations of random Debye and Einstein modes, finding very similar qualitative results.)  Again, the high-$T$ behavior is linear-in-$T$ with the low-$T$ behavior manifesting a complicated crossover.  The important point to emphasize is that the universal linear-in-$T$ resistivity always manifests for high-$T$ although the low-$T$ behavior, where the phonon-induced resistivity is generally small, shows complex crossover arising from the contributions of individual phonon modes. What is surprising and not mentioned in any early literature is that having many phonon modes does not segment this high-$T$ linearity into many different linear-$T$ regimes with different slopes, which of course will manifest as a nonlinear behavior.  No matter how many different phonon modes with different dispersions/couplings we use, we find a clear linear-in-$T$ behavior except in some fine-tuned situations.  This is significant because cuprates have many phonon modes ($\gg 10$), and therefore, one cannot argue, based just on the existence of many phonon branches in the system, that the high-$T$ linearity will be suppressed. We also mention that the realistic situation is that only $3$ of the $N$ phonon modes in any system are acoustic with $N-3$ being optical, but for our strictly theoretical analysis, this is not relevant as we use completely random phonon modes and electron-phonon couplings for all the calculations.

One may try to quantitatively define the crossover temperature $T^{*}$ that separates the low-$T$ (non-linear) and high-$T$ (linear) behavior. In the second column of Figs.~\ref{fig:Debye}--\ref{fig:Einstein}, we plot $d(\log \rho)/d(\log T)$, which extracts the exponent of $T$ in $\rho$. We see that the crossover regime from the low-$T$ behavior ($=5$ for Debye modes) to the high-$T$ behavior ($=1$) is very wide, and the behavior within that regime is very complicated. This indicates that a canonical definition of $T^*$ is impossible. Instead, we visually define $T^{*}$ as the temperature at which $d\rho/dT$ reaches about $93\%$ of the asymptotic slope ($\propto\sum_{i}\lambda_{i}$). The factor is chosen such that $T^{*}$ for a single Debye mode is exactly $T_{D}/5$. The value of $T^{*}$ for each case is shown as vertical dashed lines in Figs.~\ref{fig:Debye}--\ref{fig:Einstein}. In the third column of these figures where we plot $\rho$ as a function of $T$ in the linear scale, $T^*$ agrees very well with what appears visually to separate the nonlinear part and the linear part of the curve. In the second column of these figures, however, our choice of $T^*$ appears to be a somewhat arbitrary point within the wide crossover regime. This means that a quantitative analysis of the dependence of $T^*$ on the phonon parameters may not be meaningful.

We may still seek some rough and qualitative relationship between our choice of $T^*$ with the phonon parameters, although a precise quantitative analysis is beyond the scope of the current work as explained above. Note that $T^{*}$ is very roughly approximated by the largest $T_{D/E,i}/5$ among the individual phonon modes, since for larger $T$ all modes are linear, and for smaller $T$ at least one mode is non-linear. However, if some mode $j$ with smaller $T_{D/E,j}/5$ has $\lambda_{j}$ significantly larger than $\lambda_{i}$ (or the combined effect of several modes at that temperature scale), then $T^{*}$ may become closer to $T_{D/E,j}/5$ because the non-linearity between the two temperature scales caused by mode $i$ may be too small visually.
In fact, we do not find any simple way of expressing $T^*$ in terms of system parameters because of its complex dependence on the whole collection of $\lambda_n$ and $T_{D/E,n}$ values in a nontrivial manner.  The only concrete statement is that there is a clear linear-in-$T$ resistivity above some effective $T=T^*$, whose value depends on all the complicated details of the system, and for $T<T^*$ the resistivity does not manifest any linear-in-$T$ behavior at all.
It may be interesting in future work to investigate how $T^*$ correlates with the number of phonon modes and the statistical distributions of $T_D$ (or $T_E$) and $\lambda$. However, the ambiguity in the definition of $T^*$ makes any statistical analysis of $T^*$ ill-defined. Moreover, instead of using a statistical distribution on phonon parameters, $T^*$ is better calculated for specific materials using the specific system parameters---all our work guarantees is that there will always be a high-$T$ linear-in-$T$ resistivity independent of all the details.  This also suggests that our finding should apply to other cuprates too since all cuprates being multi-atomic materials have many phonon modes.  This problem also prevents a general statement on an effective phonon frequency and an effective electron-phonon coupling for the whole system including all the phonon modes since all the details matter, and our general theory can only establish the linearity, and not the quantitative details.  We leave these questions to future work, which would necessarily be completely numerical focusing on specific materials with many phonon modes.

One relevant question is whether there are situations where one can see several distinct well-separated linear-in-$T$ regimes with different resistivity slopes in different $T$ regimes, arising from distinct contributions of individual phonon modes.  Although we do not find such situations generically for any of our many phonon resistivity simulations (Figs.~\ref{fig:Debye}--\ref{fig:Einstein}), it is possible to create such situations through the fine-tuning of phonon parameters, particularly when the system has only a few phonon modes.  For example, consider a situation where there are two phonon modes with $\lambda_1\lesssim\lambda_2$ as well as $T_{D,1}\ll T_{D,2}$.  In this case, mode 1 produces a linear-in-$T$ regime for $T_{D,1}/5 <T$ where mode 2 contribution is still strongly suppressed.  Once $T$ reaches $T_{D,2}/5 \gg T_{D,1}/5$, mode 2 becomes quantitatively important, increasing the slope of the linear-in-$T$ resistivity.   So, such a fine-tuned situation should exhibit two separate linear-in-$T$ resistivity regimes, one for $T_{D,1}/5 < T < T_{D,2}/5$ and the other for $T>T_{D,2}/5$, and they would manifest very different resistivity slopes with the high-$T$ slope for $T>T_{D,2}$ being larger than the one in the intermediate $T_{D,1}<T<T_{D,2}$ regime.  We show such a situation in Fig.~\ref{fig:artificial} where two linear-in-$T$ segments are visually apparent as two phonon modes with fine-tuned parameters become operational in different regimes, showing an obvious change in the slope. We find this fine-tuned situation to be generically absent once many phonon modes are operational (as is the case for cuprates where typically $>20$ phonon modes contribute to scattering).  The two-mode fine-tuned situation shown in Fig.~\ref{fig:artificial} does, however, arise in simpler systems with just two phonon branches.  A well-known example is doped GaAs, where acoustic phonons produce a linear-in-$T$ regime for $T<60\,\text{K}$ and then optical phonons kick in and lead to a linear-in-$T$ regime at room temperatures whereas the intermediate temperature regime manifest nonlinear (in $T$) resistivity~\cite{Kawamura1992}.

We have also calculated the resistivity (not shown) assuming the phonon parameters to be nonrandom---for example, regularly spaced phonon coupling and frequency, and these somewhat unphysical phonon models produce results qualitatively similar to those shown in Figs.~\ref{fig:Debye}--\ref{fig:Einstein} as long as many ($\gg1$) phonon modes are considered.  The essential finding is that once many phonon modes are operational, the high-$T$ resistivity is linear in $T$ with a constant slope, and without fine-tuning, a situation with different segments manifesting different linear-in-$T$ regimes with different slopes is generally unlikely.

Our finding described above suggests that an electronic system could manifest persistent linear-in-$T$ resistivity with a constant slope above some characteristic temperature even in the presence of multiple phonon modes.  This is the generic behavior in the presence of many phonon modes, as occurring, for example, in many-atom unit cells such as high-$T_c$ cuprates (e.g.\ LSCO has 21 distinct phonon modes).  But, this finding by itself does not establish that the often-observed linear-in-$T$ resistivity in cuprates (e.g.\ optimally doped LSCO where a linear-in-$T$ resistivity persists between $60\,\text{K}$ and $1000\,\text{K}$ with essentially a constant slope) arises from phonon scattering.  All we have shown is that, as a matter of principle, phonons cannot be ruled out as the mechanism for linear-in-$T$ resistivity behavior simply because the system has many phonon modes with distinct dispersions and couplings.  To show that phonons are indeed causing the $T$-linear resistivity in a material what is necessary is a detailed quantitative transport calculation using the appropriate electronic band structure and realistic electron-phonon interactions.  This is way beyond the scope of the current work, and it is unclear that such a quantitatively reliable theory is feasible in cuprates at all, given the generally unknown electron-phonon coupling values and the unreliability of standard band theories in describing cuprates.  In particular, there is no consensus on the electron-phonon coupling constant in cuprates---in LSCO and YBCO, effective $\lambda$ values for the electron-phonon coupling are quoted, based on different estimates from theories and experiments, to be between 0.3 and 3.0~\cite{Pickett1993,Mazin1992,Zhao1994,Song1995,Cohen1990,Gadermaier2010,Benjamin2024,Yan2023,Andersen1991,Rodriguez1990,Krakauer1993}.  Instead of making any decisive claims for the cuprate resistivity, we provide below some suggestive speculations for future work by focusing on optimally doped LSCO, which manifests the most spectacular linear-in-$T$ resistivity perhaps ever seen in any metallic system~\cite{Takagi1992}.

\section{Comparison between LSCO and \texorpdfstring{$\text{Cu}$}{Cu}}

We compare the experimentally measured high-$T$ resistivity of optimally doped LSCO as reported in Ref.~\cite{Takagi1992} with the high-$T$ resistivity of one of the most studied metals: Cu.  Copper manifests a $T$-linear resistivity from $T\sim50\,\text{K}$ all the way to very high $T$, and this linearity arises entirely from the electron-phonon interaction~\cite{Matula1979}.  In Cu, $T_D \sim 340\,\text{K}$ and $\lambda \sim 0.15$, implying that Cu manifests a phonon scattering induced linear-in-$T$ resistivity starting roughly around $T \sim 50\,\text{K}$ with the resistivity slope determined by the $\lambda$ value of 0.15 (and the electronic parameters $n$ and $m$ for Cu).  For optimal doping ($x \sim 0.15$), $\text{La}_{2-x}\text{Sr}_x\text{CuO}_4$ (``LSCO'') single crystals manifest a linear-in-$T$ resistivity for $50\,\text{K}<T<1000\,\text{K}$ with the typical resistivity at $300\,\text{K}$ being $0.3\,\text{m}\Omega\,\text{cm}$ to be contrasted with the Cu resistivity at $300\,\text{K}$ being $1.5\,\mu\Omega\,\text{cm}$, making LSCO at optimal doping roughly 200 times more resistive than Cu in the $50$--$1000\,\text{K}$ regime with both the resistivity exhibiting a linear-in-$T$ behavior.

The question we ask is whether the LSCO linear-in-$T$ resistivity behavior reported first in Ref.~\cite{Takagi1992} could be crudely semiquantitatively consistent with the phonon scattering mechanism based on simple dimensional arguments analyzing the high-$T$ resistivity and comparing it with Cu.  To do this, we first use the effective Drude formula for the resistivity (following trivially from Eqs.~(\ref{eq:rhoT})--(\ref{eq:rhoBG}))
\begin{equation}\label{eq:rhoDrude}
  \rho = \frac{m}{e^2 n \tau},
\end{equation}
with the high-$T$ linear-in-$T$ form for $\tau$ following directly from Eq.~(\ref{eq:tauT}) to be:
\begin{equation}\label{eq:tauDrude}
  \frac{\hbar}{\tau} = 2\pi \lambda k_B T.
\end{equation}
We can then write the ratio of the LSCO resistivity to the Cu resistivity at the same temperature as:
\begin{equation}\label{eq:rhoLSCO}
  \frac{\rho_\text{LSCO}}{\rho_\text{Cu}} = \left(\frac{m_\text{LSCO}}{m_\text{Cu}}\right) \left(\frac{n_\text{Cu}}{n_\text{LSCO}}\right) \left(\frac{\lambda_\text{LSCO}}{\lambda_\text{Cu}}\right).
\end{equation}

For Cu, of course, the effective electron density defining the Fermi surface, $n_\text{Cu}$, and the effective conduction band carrier mass are well-established, but not so for LSCO, introducing uncertainties in our simple dimensional estimates.  It seems reasonable to assume that the LSCO has the so-called ``large'' Fermi surface at optimal doping, leading to $n_\text{LSCO}\sim n_\text{Cu}/2$ whereas the effective carrier mass for optimally doped LSCO is often taken to be $4m_e$, making $m_\text{LSCO} \sim 4 m_\text{Cu}$~\cite{Legros2019}. We note that the typical carrier effective mass in hole-doped LSCO is considered to be 3--10 $m_e$~\cite{Legros2019} and our using $m\sim 4m_e$ is a conservative choice---a larger mass would actually make our case stronger as it would further enhance the estimated LSCO resistivity, requiring even a smaller effective electron-phonon coupling.  We note that the full band structure of doped LSCO is complex (see, e.g., Ref.~\cite{Pokharel2022}) and is not important for our considerations since we are not carrying out any quantitative first-principles calculations, but are only estimating the resistivity using an operational effective mass and electron-phonon coupling. Putting these parameters in Eq.~(\ref{eq:rhoLSCO}) and using the experimental resistivity at $300\,\text{K}$ (changing this temperature obviously does not make any difference for the ratio as long as the temperature is in the linear-$T$ regime for both materials since $1/\tau \sim T$), we get the following for the ratio of the effective electron-phonon coupling constants:
\begin{equation}\label{eq:lambdaLSCO}
  \frac{\lambda_\text{LSCO}}{\lambda_\text{Cu}} = 10.
\end{equation}
Thus, an effective electron-phonon coupling of $\sim 1.5$ (i.e.\ 10 times that of $\lambda_\text{Cu}$) roughly quantitatively explains the measured LSCO resistivity at optimal doping.  Whether this relatively large coupling is reasonable or not is an open question beyond the scope of the current work, but we mention that there are many claims in the literature of $\lambda_\text{LSCO}>1$~\cite{Pickett1993,Mazin1992,Zhao1994}, and given the large number of phonon modes in LSCO, an effective $\lambda_\text{LSCO}$ for transport of $\sim 1.5$ may not be entirely unreasonable. We mention that the effective coupling here should be a suitable transport average of the couplings to all the phonon modes contributing to the carrier scattering, thus possibly enhancing its apparent strength.  In terms of the McMillan function $\alpha^2 F(\omega)$, the effective $\lambda$ is given by~\cite{McMillan1968}:
\begin{equation}\label{eq:lambdaEff}
  \lambda=2\int_0^\infty \frac{d\omega}{\omega}\alpha^2 F(\omega),
\end{equation}
where all phonon modes contribute to the integral through the $F(\omega)$ function.

We can derive an alternative equivalent expression for the resistivity in terms of the hole doping strength $x$ ($>0.1$) using the same analysis as above, obtaining:
\begin{equation}\label{eq:rhoLSCO1px}
  \rho_\text{LSCO} \sim \frac{0.7\,T}{1+x}\,\mu\Omega\,\text{cm},
\end{equation}
where $T$ is the temperature in kelvins ($>50\,\text{K}$), where the linear-in-$T$ resistivity constraint applies.  For hole doping $x <0.1$, the Fermi surface shape changes (from $1+x$ carriers per unit cell to $x$ carriers per unit cell), and Eq.~(\ref{eq:rhoLSCO1px}) is replaced by:
\begin{equation}\label{eq:rhoLSCOx}
  \rho_\text{LSCO} \sim \frac{0.7\,T}{x}\,\mu\Omega\,\text{cm}.
\end{equation}
For $x=0.2$ and $T=300\,\text{K}$, Eq.~(\ref{eq:rhoLSCO1px}) gives, $\rho \sim 0.17\,\text{m}\Omega\,\text{cm}$ whereas for $x=0.05$, Eq.~(\ref{eq:rhoLSCOx}) gives for $T=300\,\text{K}$, $\rho \sim4.5\,\text{m}\Omega\,\text{cm}$.  Both of these resistivity values are consistent with experimental measurements~\cite{Takagi1992}, where the measured resistivity~\cite{Takagi1992}  at $300\,\text{K}$ is $0.18\,\text{m}\Omega\,\text{cm}$ and $3\,\text{m}\Omega\,\text{cm}$, respectively for $x=0.2$ and $0.05$..
Equations (\ref{eq:rhoLSCO1px}) and (\ref{eq:rhoLSCOx}) simply follow from Eqs.~(\ref{eq:rhoDrude}) and (\ref{eq:tauDrude}) using $\lambda=1.5$ and the appropriate carrier density associated with the doping $x$.

Before concluding, we comment on the calculated (and measured) large resistivity values for LSCO, far surpassing the so-called Ioffe-Regel-Mott (IRM) criterion for metallicity, which for regular metals stipulates that the maximum possible metallic resistivity is $\sim 150 \,\mu\Omega\,\text{cm}$ because the metallic mean free path for more resistive samples becomes shorter than $1/k_F$ (or the lattice constant, whichever is larger).  The IRM criterion sets the limit on coherent metallic transport where the concept of momentum is still meaningful, and the claim is that a resistivity above the IRM limit represents a ``bad metal'' or an Anderson localized insulator~\cite{Gurvitch1983,Emery1995,Gunnarsson2003,Hussey2004}. In fact, the resistivity of Cu (or other metals) would reach $\sim 150 \,\mu\Omega\,\text{cm}$ for $T \sim 30\text{,}000\,\text{K}$, well above the melting temperature, and indeed regular conducting metals have a resistivity always much smaller than the IRM limit of $\sim 150 \,\mu\Omega\,\text{cm}$.  But, as is obvious here, the experimental resistivity for LSCO (and other cuprates also) typically surpasses $\sim 150 \,\mu\Omega\,\text{cm}$, making them bad metal candidates.  This is, however, easily understandable as arising, not necessarily from some exotic bad metallicity, but simply from the peculiar electronic and lattice structure of doped cuprates, which typically have much lower $k_F$ values than typical metals (and much larger lattice constant along the $c$ axis).  For example, considering the 2D limit of in-plane (in the $ab$ plane) transport, the IRM criterion of $k_F l$=1 (where $l$ is the carrier mean free path) in 2D becomes equivalent to:
\begin{equation}\label{eq:rho2DIRM}
  \rho_\text{2D IRM}=h/e^2.
\end{equation}
We convert the 2D IRM limit of Eq.~(\ref{eq:rho2DIRM}) to a corresponding 3D IRM resistivity (for a layered 3D material made out of parallel 2D layers as the cuprates are) by simply multiplying by the unit cell size along the $c$ axis ($\sim 1.3\,\text{nm}$), obtaining:
\begin{equation}\label{eq:rhoIRMLSCO}
  \rho_\text{IRM LSCO} \sim \left(\frac{h}{e^2}\right)\cdot 1.3\,\text{nm}= 3.4\,\text{m}\Omega\,\text{cm}.
\end{equation}
It does appear that the LSCO resistivity remains below this $3.4\,\text{m}\Omega\,\text{cm}$ limit even if it surpasses the $0.15\,\text{m}\Omega\,\text{cm}$ limit of normal metals. (We emphasize that Eq.~(\ref{eq:rhoIRMLSCO}) is meaningful only a 3D layered material made of 2D layers, and not for a true 3D system.) An equivalent 3D derivation of the IRM resistivity for LSCO (and other cuprates) would be simply to scale the metallic IRM limit of $0.15\,\text{m}\Omega\,\text{cm}$ by the carrier density ratio of LSCO to normal metals, giving:
\begin{equation}\label{eq:rhoIRMLSCOp}
  \rho_\text{IRM LSCO} \sim \left(\frac{350}{p}\right) \,\mu\Omega\,\text{cm},
\end{equation}
where $p=x^{1/3}$ or $(1+x)^{1/3}$ depending on whether $x <0.1$ or $>0.1$ (which decides whether the Fermi surface is ``small'' or ``large'').  Eq.~(\ref{eq:rhoIRMLSCOp}) gives a resistivity limit of  $\sim 1\,\text{m}\Omega\,\text{cm}$ for $x=0.05$ and $0.3\,\text{m}\Omega\,\text{cm}$ for $x=0.15$, which agrees semiquantitatively with the experimental data.  Obviously, the assumption of the LSCO 3D Fermi wavevector $k_F \sim x^{1/3}$ or $(1+x)^{1/3}$ is an extremely crude approximation (and therefore, an exact agreement with the experiment is not expected).  But the important point is that both Eqs.~(\ref{eq:rhoIRMLSCO}) and (\ref{eq:rhoIRMLSCOp}) imply bad metallicity for LSCO with the resistivity above the nominal metallic IRM limit of $150\,\mu\Omega\,\text{cm}$, but there is nothing profound about this violation of the metallic IRM limit as it arises from the 2D layered nature of LSCO and/or the low carrier density compared with metallic systems.
Interestingly, very similar conclusions about the IRM limit have also been established in the literature for electron-doped cuprates~\cite{Poniatowski2021}.
We mention that Refs.~\cite{Werman2016,Werman2017} consider, using the so-called ``large-$N$'' approximation, the interesting possibility that the semiclassical Boltzmann theory breaks down near the IRM limit for strong electron-phonon coupling, leading to a ``resistivity saturation''-type phenomenon in cuprates at high temperatures, where the resistivity becomes sublinear in $T$ at very high $T$ instead of increasing indefinitely linearly in $T$~\cite{Gurvitch1983,Gunnarsson2003,Millis1999}.  This would then imply that optimally doped LSCO would reflect resistivity saturation at some high $T$, ruling out bad metallicity.  There is some indirect experimental evidence~\cite{Sundqvist1990} for such a behavior if the measured resistivity (at constant pressure) is converted to the constant-volume resistivity (which is the theoretically appropriate quantity~\cite{Sundqvist1990}).
Further discussions of the actual cuprate resistivity values are beyond the scope of the current work where our focus is only on the qualitative role of having many phonon modes in the system, and not a quantitative transport theory.

We mention that although we specifically discuss optimally doped LSCO simply because this is where the most spectacular linear-in-$T$ resistivity has been experimentally reported over a large temperature range, all our considerations remain valid for other cuprates (e.g.\ YBCO) or for any material with many phonon modes.  One could wonder whether our consideration applies away from optimal doping too, and in principle, it does, but the cuprate band structure is extremely complex away from optimal doping and shows localization at low doping.  Therefore, it will be a stretch to apply our theory dar away from optimal doping because our theory is the theory for a good metal, and cuprates simply are not simple good metals away from optimal doping.  They become localized at low doping, and manifest a $\sim T^2$ resistivity at high doping, clearly indicating the appearance of competing physics arising from strong electron-electron correlations.

\section{Conclusion}

We conclude by summarizing our results and emphasizing the caveats.  Our main finding is that having many phonon modes in a system does not necessarily lead to a high-$T$ electronic resistivity which is nonlinear (and made of many linear segments with different slopes).  Our second finding is that the experimental resistivity of LSCO is consistent with the Ioffe-Regel-Mott constraint for the LSCO materials parameters, and there is nothing strange or exotic about the LSCO resistivity being larger than the nominal $150\,\mu\Omega\,\text{cm}$ IRM limit for normal metals.
Both of these findings are suggestive, and all quantitative details would depend on the materials parameters which are generally not known in cuprates.

One aspect of physics we did not discuss at all is the so-called Planckian behavior~\cite{Greene2020,Hartnoll2022,Phillips2022,Zaanen2019,Hwang2019} where the transport scattering rate $\hbar/\tau$ becomes equal to or larger than $k_B T$ because of the large resistivity of the system.  For electron-phonon scattering-limited resistivity, the Planckian behavior happens trivially whenever $\lambda > 0.16$ simply by virtue of the factor of $2\pi$ in Eq.~(\ref{eq:tauDrude}).  Thus, most normal metals manifest Planckian behavior at room temperatures since the electron-phonon scattering rate in most normal metals exceeds $k_B T$ in the linear-in-$T$ equipartition regime.  This is not a mystery or anything profound whatsoever since the scattering in this equipartition regime is basically elastic. Thus, if the cuprate resistivity in the linear-in-$T$ regime arises from electron-phonon scattering, the observed Planckian behavior is guaranteed since the effective coupling strength must be greater than 0.16.  In fact, this would be true even if electron-phonon scattering is contributing only partially to the cuprate resistivity since it is unlikely that the effective coupling arising from the many phonon modes in LSCO or YBCO is less than 0.16.  Therefore, one simple consequence of the resistive mechanism in cuprates (even partially) being electron-phonon interaction is an immediate physical explanation for the observed Planckian behavior.  If the applicable $k_F$ values are small (ie small Fermi surface), then this linear-in-$T$ strange metallicity along with the Planckian behavior would persist to $T_{BG}/5$, which for small $k_F$ could be low since $T_{BG} \sim k_F$. Whether this actually happens in cuprates or not because of electron-phonon scattering is beyond the scope of the current work.

The extent to which our findings in this work apply to realistically explaining LSCO transport properties remains an open question, mainly because the actual materials parameters for electron-phonon interaction for LSCO are not known accurately, and we ignore all scattering mechanisms other than phonon scattering at high temperatures. Moreover, from the experimental data it is very difficult to identify the crossover regime of the exponent of $T$ in $\rho$, as it is very sensitive to noise in the data. In addition, there is really no strict absolutely linear-in-$T$ resistivity anywhere even as a matter of principle, with $\rho(T)$ crossing over from a high power in $T$ (4 or 5) for $T\ll T_D$ to a linear power for $T\gg T_D$ with the crossover happening over a temperature range which depends on all the details as is obvious from our numerical results.  The same is often true for the experimental cuprate data with the resistivity only being approximately linear at best at higher temperatures as in the theory. Our work is at best suggestive, providing an incentive for further investigations of these questions using realistic (which would necessarily be numerically intensive)  models of electron-phonon coupling in cuprates.

Finally, we mention that the subject of strange metallicity and linear-in-$T$ resistivity in cuprates (or more generally, in strongly correlated materials) is a highly active research area with many publications discussing the possibility of the linear-in-$T$ resistivity (and Planckian behavior) arising from electron-electron interaction effects~\cite{Greene2020,Hartnoll2022,Phillips2022,Zaanen2019}. Some concrete examples of such theories (which should be construed as a representative, and by no means, an exhaustive reference list on the topic) can be found in Refs.~\cite{Patel2023,Huang2019,Patel2019,Allocca2024,Fournier2023}.  We emphasize that phonons are always present, and therefore, their contribution to the linear-in-$T$ resistivity is ever present even when phonon scattering may not be the dominant scattering mechanism.  An interesting unexplored question beyond the scope of the current work in this context is then why the linear-in-$T$ slope remains constant even for $T>100\,\text{K}$ where phonons are certainly producing a linear-in-$T$ resistivity if indeed the dominant resistive scattering mechanism is electron-electron scattering.  Any complete theory of carrier transport invoking electron-electron interactions should not ignore phonon scattering for $T>50\,\text{K}$ (or so) since phonon scattering is invariably present independent of the importance of electron-electron scattering.

We emphasize that our focus is only on the linear-in-$T$ resistivity, showing that the existence of many phonon modes does not generically destroy the high-temperature linear-in-$T$ resistivity in electronic materials.  We also show that the observed linear-in-$T$ resistivity in LSCO is likely consistent with a reasonable value of the electron-phonon coupling without any anomalous violation of the Ioffe-Regel-Mott metallicity bound.  There are many other aspects of strange metallicity in cuprates~\cite{Greene2020,Hartnoll2022,Phillips2022,Zaanen2019,Legros2019} which are beyond the scope of our work, and future work should ask if some of them (e.g.\ magnetotransport and Hall resistivity) could also be theoretically explained by phonon scattering---this is beyond the scope of the current work.  Finally, electron-electron interactions are undoubtedly important in cuprates, and the effects of strong correlations on the resistivity and on other aspects of strange metallicity remain a most important open question in condensed matter physics.

\section*{Acknowledgment}
The authors are grateful to Steve Kivelson for numerous discussions.  The authors also thank Jorg Schmalion and Rick Greene for discussions.
We thank Rick Greene and Nick Poniatowski for helpful comments on the manuscript.
This work is supported by the Laboratory for Physical Sciences.

\bibliographystyle{apsrev4-2}
\bibliography{references}

\begin{thebibliography}{52}%
\makeatletter
\providecommand \@ifxundefined [1]{%
 \@ifx{#1\undefined}
}%
\providecommand \@ifnum [1]{%
 \ifnum #1\expandafter \@firstoftwo
 \else \expandafter \@secondoftwo
 \fi
}%
\providecommand \@ifx [1]{%
 \ifx #1\expandafter \@firstoftwo
 \else \expandafter \@secondoftwo
 \fi
}%
\providecommand \natexlab [1]{#1}%
\providecommand \enquote  [1]{``#1''}%
\providecommand \bibnamefont  [1]{#1}%
\providecommand \bibfnamefont [1]{#1}%
\providecommand \citenamefont [1]{#1}%
\providecommand \href@noop [0]{\@secondoftwo}%
\providecommand \href [0]{\begingroup \@sanitize@url \@href}%
\providecommand \@href[1]{\@@startlink{#1}\@@href}%
\providecommand \@@href[1]{\endgroup#1\@@endlink}%
\providecommand \@sanitize@url [0]{\catcode `\\12\catcode `\$12\catcode
  `\&12\catcode `\#12\catcode `\^12\catcode `\_12\catcode `\%12\relax}%
\providecommand \@@startlink[1]{}%
\providecommand \@@endlink[0]{}%
\providecommand \url  [0]{\begingroup\@sanitize@url \@url }%
\providecommand \@url [1]{\endgroup\@href {#1}{\urlprefix }}%
\providecommand \urlprefix  [0]{URL }%
\providecommand \Eprint [0]{\href }%
\providecommand \doibase [0]{https://doi.org/}%
\providecommand \selectlanguage [0]{\@gobble}%
\providecommand \bibinfo  [0]{\@secondoftwo}%
\providecommand \bibfield  [0]{\@secondoftwo}%
\providecommand \translation [1]{[#1]}%
\providecommand \BibitemOpen [0]{}%
\providecommand \bibitemStop [0]{}%
\providecommand \bibitemNoStop [0]{.\EOS\space}%
\providecommand \EOS [0]{\spacefactor3000\relax}%
\providecommand \BibitemShut  [1]{\csname bibitem#1\endcsname}%
\let\auto@bib@innerbib\@empty
\bibitem [{\citenamefont {Greene}\ \emph {et~al.}(2020)\citenamefont {Greene},
  \citenamefont {Mandal}, \citenamefont {Poniatowski},\ and\ \citenamefont
  {Sarkar}}]{Greene2020}%
  \BibitemOpen
  \bibfield  {author} {\bibinfo {author} {\bibfnamefont {R.~L.}\ \bibnamefont
  {Greene}}, \bibinfo {author} {\bibfnamefont {P.~R.}\ \bibnamefont {Mandal}},
  \bibinfo {author} {\bibfnamefont {N.~R.}\ \bibnamefont {Poniatowski}},\ and\
  \bibinfo {author} {\bibfnamefont {T.}~\bibnamefont {Sarkar}},\ }\href
  {https://doi.org/10.1146/annurev-conmatphys-031119-050558} {\bibfield
  {journal} {\bibinfo  {journal} {Annual Review of Condensed Matter Physics}\
  }\textbf {\bibinfo {volume} {11}},\ \bibinfo {pages} {213} (\bibinfo {year}
  {2020})}\BibitemShut {NoStop}%
\bibitem [{\citenamefont {Hartnoll}\ and\ \citenamefont
  {Mackenzie}(2022)}]{Hartnoll2022}%
  \BibitemOpen
  \bibfield  {author} {\bibinfo {author} {\bibfnamefont {S.~A.}\ \bibnamefont
  {Hartnoll}}\ and\ \bibinfo {author} {\bibfnamefont {A.~P.}\ \bibnamefont
  {Mackenzie}},\ }\href {https://doi.org/10.1103/RevModPhys.94.041002}
  {\bibfield  {journal} {\bibinfo  {journal} {Rev. Mod. Phys.}\ }\textbf
  {\bibinfo {volume} {94}},\ \bibinfo {pages} {041002} (\bibinfo {year}
  {2022})}\BibitemShut {NoStop}%
\bibitem [{\citenamefont {Phillips}\ \emph {et~al.}(2022)\citenamefont
  {Phillips}, \citenamefont {Hussey},\ and\ \citenamefont
  {Abbamonte}}]{Phillips2022}%
  \BibitemOpen
  \bibfield  {author} {\bibinfo {author} {\bibfnamefont {P.~W.}\ \bibnamefont
  {Phillips}}, \bibinfo {author} {\bibfnamefont {N.~E.}\ \bibnamefont
  {Hussey}},\ and\ \bibinfo {author} {\bibfnamefont {P.}~\bibnamefont
  {Abbamonte}},\ }\href {https://doi.org/10.1126/science.abh4273} {\bibfield
  {journal} {\bibinfo  {journal} {Science}\ }\textbf {\bibinfo {volume}
  {377}},\ \bibinfo {pages} {eabh4273} (\bibinfo {year} {2022})}\BibitemShut
  {NoStop}%
\bibitem [{\citenamefont {Zaanen}(2019)}]{Zaanen2019}%
  \BibitemOpen
  \bibfield  {author} {\bibinfo {author} {\bibfnamefont {J.}~\bibnamefont
  {Zaanen}},\ }\href {https://doi.org/10.21468/SciPostPhys.6.5.061} {\bibfield
  {journal} {\bibinfo  {journal} {SciPost Phys.}\ }\textbf {\bibinfo {volume}
  {6}},\ \bibinfo {pages} {061} (\bibinfo {year} {2019})}\BibitemShut {NoStop}%
\bibitem [{\citenamefont {Legros}\ \emph {et~al.}(2019)\citenamefont {Legros},
  \citenamefont {Benhabib}, \citenamefont {Tabis}, \citenamefont
  {Lalibert{\'e}}, \citenamefont {Dion}, \citenamefont {Lizaire}, \citenamefont
  {Vignolle}, \citenamefont {Vignolles}, \citenamefont {Raffy}, \citenamefont
  {Li}, \citenamefont {Auban-Senzier}, \citenamefont {Doiron-Leyraud},
  \citenamefont {Fournier}, \citenamefont {Colson}, \citenamefont {Taillefer},\
  and\ \citenamefont {Proust}}]{Legros2019}%
  \BibitemOpen
  \bibfield  {author} {\bibinfo {author} {\bibfnamefont {A.}~\bibnamefont
  {Legros}}, \bibinfo {author} {\bibfnamefont {S.}~\bibnamefont {Benhabib}},
  \bibinfo {author} {\bibfnamefont {W.}~\bibnamefont {Tabis}}, \bibinfo
  {author} {\bibfnamefont {F.}~\bibnamefont {Lalibert{\'e}}}, \bibinfo {author}
  {\bibfnamefont {M.}~\bibnamefont {Dion}}, \bibinfo {author} {\bibfnamefont
  {M.}~\bibnamefont {Lizaire}}, \bibinfo {author} {\bibfnamefont
  {B.}~\bibnamefont {Vignolle}}, \bibinfo {author} {\bibfnamefont
  {D.}~\bibnamefont {Vignolles}}, \bibinfo {author} {\bibfnamefont
  {H.}~\bibnamefont {Raffy}}, \bibinfo {author} {\bibfnamefont {Z.~Z.}\
  \bibnamefont {Li}}, \bibinfo {author} {\bibfnamefont {P.}~\bibnamefont
  {Auban-Senzier}}, \bibinfo {author} {\bibfnamefont {N.}~\bibnamefont
  {Doiron-Leyraud}}, \bibinfo {author} {\bibfnamefont {P.}~\bibnamefont
  {Fournier}}, \bibinfo {author} {\bibfnamefont {D.}~\bibnamefont {Colson}},
  \bibinfo {author} {\bibfnamefont {L.}~\bibnamefont {Taillefer}},\ and\
  \bibinfo {author} {\bibfnamefont {C.}~\bibnamefont {Proust}},\ }\href
  {https://doi.org/10.1038/s41567-018-0334-2} {\bibfield  {journal} {\bibinfo
  {journal} {Nature Physics}\ }\textbf {\bibinfo {volume} {15}},\ \bibinfo
  {pages} {142} (\bibinfo {year} {2019})}\BibitemShut {NoStop}%
\bibitem [{\citenamefont {Takagi}\ \emph {et~al.}(1992)\citenamefont {Takagi},
  \citenamefont {Batlogg}, \citenamefont {Kao}, \citenamefont {Kwo},
  \citenamefont {Cava}, \citenamefont {Krajewski},\ and\ \citenamefont
  {Peck}}]{Takagi1992}%
  \BibitemOpen
  \bibfield  {author} {\bibinfo {author} {\bibfnamefont {H.}~\bibnamefont
  {Takagi}}, \bibinfo {author} {\bibfnamefont {B.}~\bibnamefont {Batlogg}},
  \bibinfo {author} {\bibfnamefont {H.~L.}\ \bibnamefont {Kao}}, \bibinfo
  {author} {\bibfnamefont {J.}~\bibnamefont {Kwo}}, \bibinfo {author}
  {\bibfnamefont {R.~J.}\ \bibnamefont {Cava}}, \bibinfo {author}
  {\bibfnamefont {J.~J.}\ \bibnamefont {Krajewski}},\ and\ \bibinfo {author}
  {\bibfnamefont {W.~F.}\ \bibnamefont {Peck}},\ }\href
  {https://doi.org/10.1103/PhysRevLett.69.2975} {\bibfield  {journal} {\bibinfo
   {journal} {Phys. Rev. Lett.}\ }\textbf {\bibinfo {volume} {69}},\ \bibinfo
  {pages} {2975} (\bibinfo {year} {1992})}\BibitemShut {NoStop}%
\bibitem [{\citenamefont {Komiya}\ \emph {et~al.}(2005)\citenamefont {Komiya},
  \citenamefont {Chen}, \citenamefont {Zhang},\ and\ \citenamefont
  {Ando}}]{Komiya2005}%
  \BibitemOpen
  \bibfield  {author} {\bibinfo {author} {\bibfnamefont {S.}~\bibnamefont
  {Komiya}}, \bibinfo {author} {\bibfnamefont {H.-D.}\ \bibnamefont {Chen}},
  \bibinfo {author} {\bibfnamefont {S.-C.}\ \bibnamefont {Zhang}},\ and\
  \bibinfo {author} {\bibfnamefont {Y.}~\bibnamefont {Ando}},\ }\href
  {https://doi.org/10.1103/PhysRevLett.94.207004} {\bibfield  {journal}
  {\bibinfo  {journal} {Phys. Rev. Lett.}\ }\textbf {\bibinfo {volume} {94}},\
  \bibinfo {pages} {207004} (\bibinfo {year} {2005})}\BibitemShut {NoStop}%
\bibitem [{\citenamefont {Wu}\ \emph {et~al.}(2019)\citenamefont {Wu},
  \citenamefont {Hwang},\ and\ \citenamefont {Das~Sarma}}]{Wu2019}%
  \BibitemOpen
  \bibfield  {author} {\bibinfo {author} {\bibfnamefont {F.}~\bibnamefont
  {Wu}}, \bibinfo {author} {\bibfnamefont {E.}~\bibnamefont {Hwang}},\ and\
  \bibinfo {author} {\bibfnamefont {S.}~\bibnamefont {Das~Sarma}},\ }\href
  {https://doi.org/10.1103/PhysRevB.99.165112} {\bibfield  {journal} {\bibinfo
  {journal} {Phys. Rev. B}\ }\textbf {\bibinfo {volume} {99}},\ \bibinfo
  {pages} {165112} (\bibinfo {year} {2019})}\BibitemShut {NoStop}%
\bibitem [{\citenamefont {{Das Sarma}}\ and\ \citenamefont
  {Wu}(2020)}]{DasSarma2020}%
  \BibitemOpen
  \bibfield  {author} {\bibinfo {author} {\bibfnamefont {S.}~\bibnamefont {{Das
  Sarma}}}\ and\ \bibinfo {author} {\bibfnamefont {F.}~\bibnamefont {Wu}},\
  }\href {https://doi.org/https://doi.org/10.1016/j.aop.2020.168193} {\bibfield
   {journal} {\bibinfo  {journal} {Annals of Physics}\ }\textbf {\bibinfo
  {volume} {417}},\ \bibinfo {pages} {168193} (\bibinfo {year}
  {2020})}\BibitemShut {NoStop}%
\bibitem [{\citenamefont {Hwang}\ and\ \citenamefont
  {Das~Sarma}(2019)}]{Hwang2019}%
  \BibitemOpen
  \bibfield  {author} {\bibinfo {author} {\bibfnamefont {E.~H.}\ \bibnamefont
  {Hwang}}\ and\ \bibinfo {author} {\bibfnamefont {S.}~\bibnamefont
  {Das~Sarma}},\ }\href {https://doi.org/10.1103/PhysRevB.99.085105} {\bibfield
   {journal} {\bibinfo  {journal} {Phys. Rev. B}\ }\textbf {\bibinfo {volume}
  {99}},\ \bibinfo {pages} {085105} (\bibinfo {year} {2019})}\BibitemShut
  {NoStop}%
\bibitem [{\citenamefont {Das~Sarma}\ and\ \citenamefont
  {Wu}(2022)}]{DasSarma2022}%
  \BibitemOpen
  \bibfield  {author} {\bibinfo {author} {\bibfnamefont {S.}~\bibnamefont
  {Das~Sarma}}\ and\ \bibinfo {author} {\bibfnamefont {F.}~\bibnamefont {Wu}},\
  }\href {https://doi.org/10.1103/PhysRevResearch.4.033061} {\bibfield
  {journal} {\bibinfo  {journal} {Phys. Rev. Res.}\ }\textbf {\bibinfo {volume}
  {4}},\ \bibinfo {pages} {033061} (\bibinfo {year} {2022})}\BibitemShut
  {NoStop}%
\bibitem [{\citenamefont {Sarkar}\ \emph {et~al.}(2018)\citenamefont {Sarkar},
  \citenamefont {Greene},\ and\ \citenamefont {Das~Sarma}}]{Sarkar2018}%
  \BibitemOpen
  \bibfield  {author} {\bibinfo {author} {\bibfnamefont {T.}~\bibnamefont
  {Sarkar}}, \bibinfo {author} {\bibfnamefont {R.~L.}\ \bibnamefont {Greene}},\
  and\ \bibinfo {author} {\bibfnamefont {S.}~\bibnamefont {Das~Sarma}},\ }\href
  {https://doi.org/10.1103/PhysRevB.98.224503} {\bibfield  {journal} {\bibinfo
  {journal} {Phys. Rev. B}\ }\textbf {\bibinfo {volume} {98}},\ \bibinfo
  {pages} {224503} (\bibinfo {year} {2018})}\BibitemShut {NoStop}%
\bibitem [{\citenamefont {Ziman}(2001)}]{Ziman2001}%
  \BibitemOpen
  \bibfield  {author} {\bibinfo {author} {\bibfnamefont {J.}~\bibnamefont
  {Ziman}},\ }\href {https://doi.org/10.1093/acprof:oso/9780198507796.001.0001}
  {\emph {\bibinfo {title} {{Electrons and Phonons: The Theory of Transport
  Phenomena in Solids}}}}\ (\bibinfo  {publisher} {Oxford University Press},\
  \bibinfo {year} {2001})\BibitemShut {NoStop}%
\bibitem [{\citenamefont {Ashcroft}\ and\ \citenamefont
  {Mermin}(1976)}]{Ashcroft1976}%
  \BibitemOpen
  \bibfield  {author} {\bibinfo {author} {\bibfnamefont {N.~W.}\ \bibnamefont
  {Ashcroft}}\ and\ \bibinfo {author} {\bibfnamefont {N.~D.}\ \bibnamefont
  {Mermin}},\ }\href@noop {} {\emph {\bibinfo {title} {Solid State Physics}}}\
  (\bibinfo  {publisher} {Saunders College Publishing},\ \bibinfo {year}
  {1976})\BibitemShut {NoStop}%
\bibitem [{\citenamefont {Grimvall}(1976)}]{Grimvall1976}%
  \BibitemOpen
  \bibfield  {author} {\bibinfo {author} {\bibfnamefont {G.}~\bibnamefont
  {Grimvall}},\ }\href {https://doi.org/10.1088/0031-8949/14/1-2/013}
  {\bibfield  {journal} {\bibinfo  {journal} {Physica Scripta}\ }\textbf
  {\bibinfo {volume} {14}},\ \bibinfo {pages} {63} (\bibinfo {year}
  {1976})}\BibitemShut {NoStop}%
\bibitem [{\citenamefont {Min}\ \emph {et~al.}(2012)\citenamefont {Min},
  \citenamefont {Hwang},\ and\ \citenamefont {Das~Sarma}}]{Min2012}%
  \BibitemOpen
  \bibfield  {author} {\bibinfo {author} {\bibfnamefont {H.}~\bibnamefont
  {Min}}, \bibinfo {author} {\bibfnamefont {E.~H.}\ \bibnamefont {Hwang}},\
  and\ \bibinfo {author} {\bibfnamefont {S.}~\bibnamefont {Das~Sarma}},\ }\href
  {https://doi.org/10.1103/PhysRevB.86.085307} {\bibfield  {journal} {\bibinfo
  {journal} {Phys. Rev. B}\ }\textbf {\bibinfo {volume} {86}},\ \bibinfo
  {pages} {085307} (\bibinfo {year} {2012})}\BibitemShut {NoStop}%
\bibitem [{\citenamefont {Ziman}(1972)}]{Ziman1972}%
  \BibitemOpen
  \bibfield  {author} {\bibinfo {author} {\bibfnamefont {J.~M.}\ \bibnamefont
  {Ziman}},\ }\href@noop {} {\emph {\bibinfo {title} {Principles of the Theory
  of Solids}}},\ \bibinfo {edition} {2nd}\ ed.\ (\bibinfo  {publisher}
  {Cambridge University Press},\ \bibinfo {year} {1972})\BibitemShut {NoStop}%
\bibitem [{\citenamefont {Allen}(1999)}]{Allen1999}%
  \BibitemOpen
  \bibfield  {author} {\bibinfo {author} {\bibfnamefont {P.~B.}\ \bibnamefont
  {Allen}},\ }in\ \href@noop {} {\emph {\bibinfo {booktitle} {Handbook of
  Superconductivity}}},\ \bibinfo {editor} {edited by\ \bibinfo {editor}
  {\bibfnamefont {C.~P.}\ \bibnamefont {Poole}}}\ (\bibinfo  {publisher}
  {Academic Press},\ \bibinfo {address} {New York},\ \bibinfo {year} {1999})\
  Chap.~\bibinfo {chapter} {9}\BibitemShut {NoStop}%
\bibitem [{\citenamefont {Kawamura}\ and\ \citenamefont
  {Das~Sarma}(1992)}]{Kawamura1992}%
  \BibitemOpen
  \bibfield  {author} {\bibinfo {author} {\bibfnamefont {T.}~\bibnamefont
  {Kawamura}}\ and\ \bibinfo {author} {\bibfnamefont {S.}~\bibnamefont
  {Das~Sarma}},\ }\href {https://doi.org/10.1103/PhysRevB.45.3612} {\bibfield
  {journal} {\bibinfo  {journal} {Phys. Rev. B}\ }\textbf {\bibinfo {volume}
  {45}},\ \bibinfo {pages} {3612} (\bibinfo {year} {1992})}\BibitemShut
  {NoStop}%
\bibitem [{\citenamefont {Allen}(1978)}]{Allen1978}%
  \BibitemOpen
  \bibfield  {author} {\bibinfo {author} {\bibfnamefont {P.~B.}\ \bibnamefont
  {Allen}},\ }\href {https://doi.org/10.1103/PhysRevB.17.3725} {\bibfield
  {journal} {\bibinfo  {journal} {Phys. Rev. B}\ }\textbf {\bibinfo {volume}
  {17}},\ \bibinfo {pages} {3725} (\bibinfo {year} {1978})}\BibitemShut
  {NoStop}%
\bibitem [{\citenamefont {Allen}\ \emph {et~al.}(1988)\citenamefont {Allen},
  \citenamefont {Pickett},\ and\ \citenamefont {Krakauer}}]{Allen1988}%
  \BibitemOpen
  \bibfield  {author} {\bibinfo {author} {\bibfnamefont {P.~B.}\ \bibnamefont
  {Allen}}, \bibinfo {author} {\bibfnamefont {W.~E.}\ \bibnamefont {Pickett}},\
  and\ \bibinfo {author} {\bibfnamefont {H.}~\bibnamefont {Krakauer}},\ }\href
  {https://doi.org/10.1103/PhysRevB.37.7482} {\bibfield  {journal} {\bibinfo
  {journal} {Phys. Rev. B}\ }\textbf {\bibinfo {volume} {37}},\ \bibinfo
  {pages} {7482} (\bibinfo {year} {1988})}\BibitemShut {NoStop}%
\bibitem [{\citenamefont {Macheda}\ and\ \citenamefont
  {Bonini}(2018)}]{Macheda2018}%
  \BibitemOpen
  \bibfield  {author} {\bibinfo {author} {\bibfnamefont {F.}~\bibnamefont
  {Macheda}}\ and\ \bibinfo {author} {\bibfnamefont {N.}~\bibnamefont
  {Bonini}},\ }\href {https://doi.org/10.1103/PhysRevB.98.201201} {\bibfield
  {journal} {\bibinfo  {journal} {Phys. Rev. B}\ }\textbf {\bibinfo {volume}
  {98}},\ \bibinfo {pages} {201201} (\bibinfo {year} {2018})}\BibitemShut
  {NoStop}%
\bibitem [{\citenamefont {Ponc\'e}\ \emph {et~al.}(2021)\citenamefont
  {Ponc\'e}, \citenamefont {Macheda}, \citenamefont {Margine}, \citenamefont
  {Marzari}, \citenamefont {Bonini},\ and\ \citenamefont
  {Giustino}}]{Ponce2021}%
  \BibitemOpen
  \bibfield  {author} {\bibinfo {author} {\bibfnamefont {S.}~\bibnamefont
  {Ponc\'e}}, \bibinfo {author} {\bibfnamefont {F.}~\bibnamefont {Macheda}},
  \bibinfo {author} {\bibfnamefont {E.~R.}\ \bibnamefont {Margine}}, \bibinfo
  {author} {\bibfnamefont {N.}~\bibnamefont {Marzari}}, \bibinfo {author}
  {\bibfnamefont {N.}~\bibnamefont {Bonini}},\ and\ \bibinfo {author}
  {\bibfnamefont {F.}~\bibnamefont {Giustino}},\ }\href
  {https://doi.org/10.1103/PhysRevResearch.3.043022} {\bibfield  {journal}
  {\bibinfo  {journal} {Phys. Rev. Res.}\ }\textbf {\bibinfo {volume} {3}},\
  \bibinfo {pages} {043022} (\bibinfo {year} {2021})}\BibitemShut {NoStop}%
\bibitem [{\citenamefont {Davis}\ and\ \citenamefont
  {Sarma}(2023)}]{Davis2023}%
  \BibitemOpen
  \bibfield  {author} {\bibinfo {author} {\bibfnamefont {S.~M.}\ \bibnamefont
  {Davis}}\ and\ \bibinfo {author} {\bibfnamefont {S.~D.}\ \bibnamefont
  {Sarma}},\ }\href@noop {} {\bibinfo {title} {The kinetic theory of
  ultra-subsonic fermion systems and applications to flat band magic angle
  twisted bilayer graphene}} (\bibinfo {year} {2023}),\ \Eprint
  {https://arxiv.org/abs/2305.09120} {arXiv:2305.09120 [cond-mat.mes-hall]}
  \BibitemShut {NoStop}%
\bibitem [{\citenamefont {Pickett}\ \emph {et~al.}(1993)\citenamefont
  {Pickett}, \citenamefont {Krakauer},\ and\ \citenamefont
  {Cohen}}]{Pickett1993}%
  \BibitemOpen
  \bibfield  {author} {\bibinfo {author} {\bibfnamefont {W.}~\bibnamefont
  {Pickett}}, \bibinfo {author} {\bibfnamefont {H.}~\bibnamefont {Krakauer}},\
  and\ \bibinfo {author} {\bibfnamefont {R.}~\bibnamefont {Cohen}},\ }\href
  {https://doi.org/https://doi.org/10.1016/0964-1807(93)90236-U} {\bibfield
  {journal} {\bibinfo  {journal} {Applied Superconductivity}\ }\textbf
  {\bibinfo {volume} {1}},\ \bibinfo {pages} {251} (\bibinfo {year}
  {1993})}\BibitemShut {NoStop}%
\bibitem [{\citenamefont {Mazin}\ and\ \citenamefont
  {Dolgov}(1992)}]{Mazin1992}%
  \BibitemOpen
  \bibfield  {author} {\bibinfo {author} {\bibfnamefont {I.~I.}\ \bibnamefont
  {Mazin}}\ and\ \bibinfo {author} {\bibfnamefont {O.~V.}\ \bibnamefont
  {Dolgov}},\ }\href {https://doi.org/10.1103/PhysRevB.45.2509} {\bibfield
  {journal} {\bibinfo  {journal} {Phys. Rev. B}\ }\textbf {\bibinfo {volume}
  {45}},\ \bibinfo {pages} {2509} (\bibinfo {year} {1992})}\BibitemShut
  {NoStop}%
\bibitem [{\citenamefont {Zhao}\ and\ \citenamefont
  {Callaway}(1994)}]{Zhao1994}%
  \BibitemOpen
  \bibfield  {author} {\bibinfo {author} {\bibfnamefont {G.~L.}\ \bibnamefont
  {Zhao}}\ and\ \bibinfo {author} {\bibfnamefont {J.}~\bibnamefont
  {Callaway}},\ }\href {https://doi.org/10.1103/PhysRevB.50.9511} {\bibfield
  {journal} {\bibinfo  {journal} {Phys. Rev. B}\ }\textbf {\bibinfo {volume}
  {50}},\ \bibinfo {pages} {9511} (\bibinfo {year} {1994})}\BibitemShut
  {NoStop}%
\bibitem [{\citenamefont {Song}\ and\ \citenamefont {Annett}(1995)}]{Song1995}%
  \BibitemOpen
  \bibfield  {author} {\bibinfo {author} {\bibfnamefont {J.}~\bibnamefont
  {Song}}\ and\ \bibinfo {author} {\bibfnamefont {J.~F.}\ \bibnamefont
  {Annett}},\ }\href {https://doi.org/10.1103/PhysRevB.51.3840} {\bibfield
  {journal} {\bibinfo  {journal} {Phys. Rev. B}\ }\textbf {\bibinfo {volume}
  {51}},\ \bibinfo {pages} {3840} (\bibinfo {year} {1995})}\BibitemShut
  {NoStop}%
\bibitem [{\citenamefont {Cohen}\ \emph {et~al.}(1990)\citenamefont {Cohen},
  \citenamefont {Pickett},\ and\ \citenamefont {Krakauer}}]{Cohen1990}%
  \BibitemOpen
  \bibfield  {author} {\bibinfo {author} {\bibfnamefont {R.~E.}\ \bibnamefont
  {Cohen}}, \bibinfo {author} {\bibfnamefont {W.~E.}\ \bibnamefont {Pickett}},\
  and\ \bibinfo {author} {\bibfnamefont {H.}~\bibnamefont {Krakauer}},\ }\href
  {https://doi.org/10.1103/PhysRevLett.64.2575} {\bibfield  {journal} {\bibinfo
   {journal} {Phys. Rev. Lett.}\ }\textbf {\bibinfo {volume} {64}},\ \bibinfo
  {pages} {2575} (\bibinfo {year} {1990})}\BibitemShut {NoStop}%
\bibitem [{\citenamefont {Gadermaier}\ \emph {et~al.}(2010)\citenamefont
  {Gadermaier}, \citenamefont {Alexandrov}, \citenamefont {Kabanov},
  \citenamefont {Kusar}, \citenamefont {Mertelj}, \citenamefont {Yao},
  \citenamefont {Manzoni}, \citenamefont {Brida}, \citenamefont {Cerullo},\
  and\ \citenamefont {Mihailovic}}]{Gadermaier2010}%
  \BibitemOpen
  \bibfield  {author} {\bibinfo {author} {\bibfnamefont {C.}~\bibnamefont
  {Gadermaier}}, \bibinfo {author} {\bibfnamefont {A.~S.}\ \bibnamefont
  {Alexandrov}}, \bibinfo {author} {\bibfnamefont {V.~V.}\ \bibnamefont
  {Kabanov}}, \bibinfo {author} {\bibfnamefont {P.}~\bibnamefont {Kusar}},
  \bibinfo {author} {\bibfnamefont {T.}~\bibnamefont {Mertelj}}, \bibinfo
  {author} {\bibfnamefont {X.}~\bibnamefont {Yao}}, \bibinfo {author}
  {\bibfnamefont {C.}~\bibnamefont {Manzoni}}, \bibinfo {author} {\bibfnamefont
  {D.}~\bibnamefont {Brida}}, \bibinfo {author} {\bibfnamefont
  {G.}~\bibnamefont {Cerullo}},\ and\ \bibinfo {author} {\bibfnamefont
  {D.}~\bibnamefont {Mihailovic}},\ }\href
  {https://doi.org/10.1103/PhysRevLett.105.257001} {\bibfield  {journal}
  {\bibinfo  {journal} {Phys. Rev. Lett.}\ }\textbf {\bibinfo {volume} {105}},\
  \bibinfo {pages} {257001} (\bibinfo {year} {2010})}\BibitemShut {NoStop}%
\bibitem [{\citenamefont {Chang}\ \emph {et~al.}(2024)\citenamefont {Chang},
  \citenamefont {Timrov}, \citenamefont {Park}, \citenamefont {Zhou},
  \citenamefont {Marzari},\ and\ \citenamefont {Bernardi}}]{Benjamin2024}%
  \BibitemOpen
  \bibfield  {author} {\bibinfo {author} {\bibfnamefont {B.~K.}\ \bibnamefont
  {Chang}}, \bibinfo {author} {\bibfnamefont {I.}~\bibnamefont {Timrov}},
  \bibinfo {author} {\bibfnamefont {J.}~\bibnamefont {Park}}, \bibinfo {author}
  {\bibfnamefont {J.-J.}\ \bibnamefont {Zhou}}, \bibinfo {author}
  {\bibfnamefont {N.}~\bibnamefont {Marzari}},\ and\ \bibinfo {author}
  {\bibfnamefont {M.}~\bibnamefont {Bernardi}},\ }\href@noop {} {} (\bibinfo
  {year} {2024}),\ \Eprint {https://arxiv.org/abs/2401.11322} {arXiv:2401.11322
  [cond-mat.mtrl-sci]} \BibitemShut {NoStop}%
\bibitem [{\citenamefont {Yan}\ \emph {et~al.}(2023)\citenamefont {Yan},
  \citenamefont {Bok}, \citenamefont {He}, \citenamefont {Zhang}, \citenamefont
  {Gao}, \citenamefont {Luo}, \citenamefont {Cai}, \citenamefont {Peng},
  \citenamefont {Meng}, \citenamefont {Li}, \citenamefont {Chen}, \citenamefont
  {Song}, \citenamefont {Yin}, \citenamefont {Miao}, \citenamefont {Chen},
  \citenamefont {Gu}, \citenamefont {Lin}, \citenamefont {Zhang}, \citenamefont
  {Yang}, \citenamefont {Zhang}, \citenamefont {Peng}, \citenamefont {Liu},
  \citenamefont {Zhao}, \citenamefont {Choi}, \citenamefont {Xu},\ and\
  \citenamefont {Zhou}}]{Yan2023}%
  \BibitemOpen
  \bibfield  {author} {\bibinfo {author} {\bibfnamefont {H.}~\bibnamefont
  {Yan}}, \bibinfo {author} {\bibfnamefont {J.~M.}\ \bibnamefont {Bok}},
  \bibinfo {author} {\bibfnamefont {J.}~\bibnamefont {He}}, \bibinfo {author}
  {\bibfnamefont {W.}~\bibnamefont {Zhang}}, \bibinfo {author} {\bibfnamefont
  {Q.}~\bibnamefont {Gao}}, \bibinfo {author} {\bibfnamefont {X.}~\bibnamefont
  {Luo}}, \bibinfo {author} {\bibfnamefont {Y.}~\bibnamefont {Cai}}, \bibinfo
  {author} {\bibfnamefont {Y.}~\bibnamefont {Peng}}, \bibinfo {author}
  {\bibfnamefont {J.}~\bibnamefont {Meng}}, \bibinfo {author} {\bibfnamefont
  {C.}~\bibnamefont {Li}}, \bibinfo {author} {\bibfnamefont {H.}~\bibnamefont
  {Chen}}, \bibinfo {author} {\bibfnamefont {C.}~\bibnamefont {Song}}, \bibinfo
  {author} {\bibfnamefont {C.}~\bibnamefont {Yin}}, \bibinfo {author}
  {\bibfnamefont {T.}~\bibnamefont {Miao}}, \bibinfo {author} {\bibfnamefont
  {Y.}~\bibnamefont {Chen}}, \bibinfo {author} {\bibfnamefont {G.}~\bibnamefont
  {Gu}}, \bibinfo {author} {\bibfnamefont {C.}~\bibnamefont {Lin}}, \bibinfo
  {author} {\bibfnamefont {F.}~\bibnamefont {Zhang}}, \bibinfo {author}
  {\bibfnamefont {F.}~\bibnamefont {Yang}}, \bibinfo {author} {\bibfnamefont
  {S.}~\bibnamefont {Zhang}}, \bibinfo {author} {\bibfnamefont
  {Q.}~\bibnamefont {Peng}}, \bibinfo {author} {\bibfnamefont {G.}~\bibnamefont
  {Liu}}, \bibinfo {author} {\bibfnamefont {L.}~\bibnamefont {Zhao}}, \bibinfo
  {author} {\bibfnamefont {H.-Y.}\ \bibnamefont {Choi}}, \bibinfo {author}
  {\bibfnamefont {Z.}~\bibnamefont {Xu}},\ and\ \bibinfo {author}
  {\bibfnamefont {X.~J.}\ \bibnamefont {Zhou}},\ }\href
  {https://doi.org/10.1073/pnas.2219491120} {\bibfield  {journal} {\bibinfo
  {journal} {Proceedings of the National Academy of Sciences}\ }\textbf
  {\bibinfo {volume} {120}},\ \bibinfo {pages} {e2219491120} (\bibinfo {year}
  {2023})}\BibitemShut {NoStop}%
\bibitem [{\citenamefont {Andersen}\ \emph {et~al.}(1991)\citenamefont
  {Andersen}, \citenamefont {Liechtenstein}, \citenamefont {Rodriguez},
  \citenamefont {Mazin}, \citenamefont {Jepsen}, \citenamefont {Antropov},
  \citenamefont {Gunnarsson},\ and\ \citenamefont {Gopalan}}]{Andersen1991}%
  \BibitemOpen
  \bibfield  {author} {\bibinfo {author} {\bibfnamefont {O.}~\bibnamefont
  {Andersen}}, \bibinfo {author} {\bibfnamefont {A.}~\bibnamefont
  {Liechtenstein}}, \bibinfo {author} {\bibfnamefont {O.}~\bibnamefont
  {Rodriguez}}, \bibinfo {author} {\bibfnamefont {I.}~\bibnamefont {Mazin}},
  \bibinfo {author} {\bibfnamefont {O.}~\bibnamefont {Jepsen}}, \bibinfo
  {author} {\bibfnamefont {V.}~\bibnamefont {Antropov}}, \bibinfo {author}
  {\bibfnamefont {O.}~\bibnamefont {Gunnarsson}},\ and\ \bibinfo {author}
  {\bibfnamefont {S.}~\bibnamefont {Gopalan}},\ }\href
  {https://doi.org/https://doi.org/10.1016/0921-4534(91)91964-6} {\bibfield
  {journal} {\bibinfo  {journal} {Physica C: Superconductivity}\ }\textbf
  {\bibinfo {volume} {185-189}},\ \bibinfo {pages} {147} (\bibinfo {year}
  {1991})}\BibitemShut {NoStop}%
\bibitem [{\citenamefont {Rodriguez}\ \emph {et~al.}(1990)\citenamefont
  {Rodriguez}, \citenamefont {Liechtenstein}, \citenamefont {Mazin},
  \citenamefont {Jepsen}, \citenamefont {Andersen},\ and\ \citenamefont
  {Methfessel}}]{Rodriguez1990}%
  \BibitemOpen
  \bibfield  {author} {\bibinfo {author} {\bibfnamefont {C.~O.}\ \bibnamefont
  {Rodriguez}}, \bibinfo {author} {\bibfnamefont {A.~I.}\ \bibnamefont
  {Liechtenstein}}, \bibinfo {author} {\bibfnamefont {I.~I.}\ \bibnamefont
  {Mazin}}, \bibinfo {author} {\bibfnamefont {O.}~\bibnamefont {Jepsen}},
  \bibinfo {author} {\bibfnamefont {O.~K.}\ \bibnamefont {Andersen}},\ and\
  \bibinfo {author} {\bibfnamefont {M.}~\bibnamefont {Methfessel}},\ }\href
  {https://doi.org/10.1103/PhysRevB.42.2692} {\bibfield  {journal} {\bibinfo
  {journal} {Phys. Rev. B}\ }\textbf {\bibinfo {volume} {42}},\ \bibinfo
  {pages} {2692} (\bibinfo {year} {1990})}\BibitemShut {NoStop}%
\bibitem [{\citenamefont {Krakauer}\ \emph {et~al.}(1993)\citenamefont
  {Krakauer}, \citenamefont {Pickett},\ and\ \citenamefont
  {Cohen}}]{Krakauer1993}%
  \BibitemOpen
  \bibfield  {author} {\bibinfo {author} {\bibfnamefont {H.}~\bibnamefont
  {Krakauer}}, \bibinfo {author} {\bibfnamefont {W.~E.}\ \bibnamefont
  {Pickett}},\ and\ \bibinfo {author} {\bibfnamefont {R.~E.}\ \bibnamefont
  {Cohen}},\ }\href {https://doi.org/10.1103/PhysRevB.47.1002} {\bibfield
  {journal} {\bibinfo  {journal} {Phys. Rev. B}\ }\textbf {\bibinfo {volume}
  {47}},\ \bibinfo {pages} {1002} (\bibinfo {year} {1993})}\BibitemShut
  {NoStop}%
\bibitem [{\citenamefont {Matula}(1979)}]{Matula1979}%
  \BibitemOpen
  \bibfield  {author} {\bibinfo {author} {\bibfnamefont {R.~A.}\ \bibnamefont
  {Matula}},\ }\href {https://doi.org/10.1063/1.555614} {\bibfield  {journal}
  {\bibinfo  {journal} {Journal of Physical and Chemical Reference Data}\
  }\textbf {\bibinfo {volume} {8}},\ \bibinfo {pages} {1147} (\bibinfo {year}
  {1979})}\BibitemShut {NoStop}%
\bibitem [{\citenamefont {Pokharel}\ \emph {et~al.}(2022)\citenamefont
  {Pokharel}, \citenamefont {Lane}, \citenamefont {Furness}, \citenamefont
  {Zhang}, \citenamefont {Ning}, \citenamefont {Barbiellini}, \citenamefont
  {Markiewicz}, \citenamefont {Zhang}, \citenamefont {Bansil},\ and\
  \citenamefont {Sun}}]{Pokharel2022}%
  \BibitemOpen
  \bibfield  {author} {\bibinfo {author} {\bibfnamefont {K.}~\bibnamefont
  {Pokharel}}, \bibinfo {author} {\bibfnamefont {C.}~\bibnamefont {Lane}},
  \bibinfo {author} {\bibfnamefont {J.~W.}\ \bibnamefont {Furness}}, \bibinfo
  {author} {\bibfnamefont {R.}~\bibnamefont {Zhang}}, \bibinfo {author}
  {\bibfnamefont {J.}~\bibnamefont {Ning}}, \bibinfo {author} {\bibfnamefont
  {B.}~\bibnamefont {Barbiellini}}, \bibinfo {author} {\bibfnamefont {R.~S.}\
  \bibnamefont {Markiewicz}}, \bibinfo {author} {\bibfnamefont
  {Y.}~\bibnamefont {Zhang}}, \bibinfo {author} {\bibfnamefont
  {A.}~\bibnamefont {Bansil}},\ and\ \bibinfo {author} {\bibfnamefont
  {J.}~\bibnamefont {Sun}},\ }\href@noop {} {\bibinfo {title} {Sensitivity of
  the electronic and magnetic structures of cuprate superconductors to density
  functional approximations}} (\bibinfo {year} {2022}),\ \Eprint
  {https://arxiv.org/abs/2004.08047} {arXiv:2004.08047 [cond-mat.supr-con]}
  \BibitemShut {NoStop}%
\bibitem [{\citenamefont {McMillan}(1968)}]{McMillan1968}%
  \BibitemOpen
  \bibfield  {author} {\bibinfo {author} {\bibfnamefont {W.~L.}\ \bibnamefont
  {McMillan}},\ }\href {https://doi.org/10.1103/PhysRev.167.331} {\bibfield
  {journal} {\bibinfo  {journal} {Phys. Rev.}\ }\textbf {\bibinfo {volume}
  {167}},\ \bibinfo {pages} {331} (\bibinfo {year} {1968})}\BibitemShut
  {NoStop}%
\bibitem [{\citenamefont {Gurvitch}(1983)}]{Gurvitch1983}%
  \BibitemOpen
  \bibfield  {author} {\bibinfo {author} {\bibfnamefont {M.}~\bibnamefont
  {Gurvitch}},\ }\href {https://doi.org/10.1103/PhysRevB.28.544} {\bibfield
  {journal} {\bibinfo  {journal} {Phys. Rev. B}\ }\textbf {\bibinfo {volume}
  {28}},\ \bibinfo {pages} {544} (\bibinfo {year} {1983})}\BibitemShut
  {NoStop}%
\bibitem [{\citenamefont {Emery}\ and\ \citenamefont
  {Kivelson}(1995)}]{Emery1995}%
  \BibitemOpen
  \bibfield  {author} {\bibinfo {author} {\bibfnamefont {V.~J.}\ \bibnamefont
  {Emery}}\ and\ \bibinfo {author} {\bibfnamefont {S.~A.}\ \bibnamefont
  {Kivelson}},\ }\href {https://doi.org/10.1103/PhysRevLett.74.3253} {\bibfield
   {journal} {\bibinfo  {journal} {Phys. Rev. Lett.}\ }\textbf {\bibinfo
  {volume} {74}},\ \bibinfo {pages} {3253} (\bibinfo {year}
  {1995})}\BibitemShut {NoStop}%
\bibitem [{\citenamefont {Gunnarsson}\ \emph {et~al.}(2003)\citenamefont
  {Gunnarsson}, \citenamefont {Calandra},\ and\ \citenamefont
  {Han}}]{Gunnarsson2003}%
  \BibitemOpen
  \bibfield  {author} {\bibinfo {author} {\bibfnamefont {O.}~\bibnamefont
  {Gunnarsson}}, \bibinfo {author} {\bibfnamefont {M.}~\bibnamefont
  {Calandra}},\ and\ \bibinfo {author} {\bibfnamefont {J.~E.}\ \bibnamefont
  {Han}},\ }\href {https://doi.org/10.1103/RevModPhys.75.1085} {\bibfield
  {journal} {\bibinfo  {journal} {Rev. Mod. Phys.}\ }\textbf {\bibinfo {volume}
  {75}},\ \bibinfo {pages} {1085} (\bibinfo {year} {2003})}\BibitemShut
  {NoStop}%
\bibitem [{\citenamefont {Hussey}\ \emph {et~al.}(2004)\citenamefont {Hussey},
  \citenamefont {Takenaka},\ and\ \citenamefont {Takagi}}]{Hussey2004}%
  \BibitemOpen
  \bibfield  {author} {\bibinfo {author} {\bibfnamefont {N.~E.}\ \bibnamefont
  {Hussey}}, \bibinfo {author} {\bibfnamefont {K.}~\bibnamefont {Takenaka}},\
  and\ \bibinfo {author} {\bibfnamefont {H.}~\bibnamefont {Takagi}},\ }\href
  {https://doi.org/10.1080/14786430410001716944} {\bibfield  {journal}
  {\bibinfo  {journal} {Philosophical Magazine}\ }\textbf {\bibinfo {volume}
  {84}},\ \bibinfo {pages} {2847} (\bibinfo {year} {2004})}\BibitemShut
  {NoStop}%
\bibitem [{\citenamefont {Poniatowski}\ \emph {et~al.}(2021)\citenamefont
  {Poniatowski}, \citenamefont {Sarkar}, \citenamefont {Das~Sarma},\ and\
  \citenamefont {Greene}}]{Poniatowski2021}%
  \BibitemOpen
  \bibfield  {author} {\bibinfo {author} {\bibfnamefont {N.~R.}\ \bibnamefont
  {Poniatowski}}, \bibinfo {author} {\bibfnamefont {T.}~\bibnamefont {Sarkar}},
  \bibinfo {author} {\bibfnamefont {S.}~\bibnamefont {Das~Sarma}},\ and\
  \bibinfo {author} {\bibfnamefont {R.~L.}\ \bibnamefont {Greene}},\ }\href
  {https://doi.org/10.1103/PhysRevB.103.L020501} {\bibfield  {journal}
  {\bibinfo  {journal} {Phys. Rev. B}\ }\textbf {\bibinfo {volume} {103}},\
  \bibinfo {pages} {L020501} (\bibinfo {year} {2021})}\BibitemShut {NoStop}%
\bibitem [{\citenamefont {Werman}\ and\ \citenamefont
  {Berg}(2016)}]{Werman2016}%
  \BibitemOpen
  \bibfield  {author} {\bibinfo {author} {\bibfnamefont {Y.}~\bibnamefont
  {Werman}}\ and\ \bibinfo {author} {\bibfnamefont {E.}~\bibnamefont {Berg}},\
  }\href {https://doi.org/10.1103/PhysRevB.93.075109} {\bibfield  {journal}
  {\bibinfo  {journal} {Phys. Rev. B}\ }\textbf {\bibinfo {volume} {93}},\
  \bibinfo {pages} {075109} (\bibinfo {year} {2016})}\BibitemShut {NoStop}%
\bibitem [{\citenamefont {Werman}\ \emph {et~al.}(2017)\citenamefont {Werman},
  \citenamefont {Kivelson},\ and\ \citenamefont {Berg}}]{Werman2017}%
  \BibitemOpen
  \bibfield  {author} {\bibinfo {author} {\bibfnamefont {Y.}~\bibnamefont
  {Werman}}, \bibinfo {author} {\bibfnamefont {S.~A.}\ \bibnamefont
  {Kivelson}},\ and\ \bibinfo {author} {\bibfnamefont {E.}~\bibnamefont
  {Berg}},\ }\href {https://doi.org/10.1038/s41535-017-0009-8} {\bibfield
  {journal} {\bibinfo  {journal} {npj Quantum Materials}\ }\textbf {\bibinfo
  {volume} {2}},\ \bibinfo {pages} {7} (\bibinfo {year} {2017})}\BibitemShut
  {NoStop}%
\bibitem [{\citenamefont {Millis}\ \emph {et~al.}(1999)\citenamefont {Millis},
  \citenamefont {Hu},\ and\ \citenamefont {Das~Sarma}}]{Millis1999}%
  \BibitemOpen
  \bibfield  {author} {\bibinfo {author} {\bibfnamefont {A.~J.}\ \bibnamefont
  {Millis}}, \bibinfo {author} {\bibfnamefont {J.}~\bibnamefont {Hu}},\ and\
  \bibinfo {author} {\bibfnamefont {S.}~\bibnamefont {Das~Sarma}},\ }\href
  {https://doi.org/10.1103/PhysRevLett.82.2354} {\bibfield  {journal} {\bibinfo
   {journal} {Phys. Rev. Lett.}\ }\textbf {\bibinfo {volume} {82}},\ \bibinfo
  {pages} {2354} (\bibinfo {year} {1999})}\BibitemShut {NoStop}%
\bibitem [{\citenamefont {Sundqvist}\ and\ \citenamefont
  {Andersson}(1990)}]{Sundqvist1990}%
  \BibitemOpen
  \bibfield  {author} {\bibinfo {author} {\bibfnamefont {B.}~\bibnamefont
  {Sundqvist}}\ and\ \bibinfo {author} {\bibfnamefont {B.}~\bibnamefont
  {Andersson}},\ }\href
  {https://doi.org/https://doi.org/10.1016/0038-1098(90)90076-N} {\bibfield
  {journal} {\bibinfo  {journal} {Solid State Communications}\ }\textbf
  {\bibinfo {volume} {76}},\ \bibinfo {pages} {1019} (\bibinfo {year}
  {1990})}\BibitemShut {NoStop}%
\bibitem [{\citenamefont {Patel}\ \emph {et~al.}(2023)\citenamefont {Patel},
  \citenamefont {Guo}, \citenamefont {Esterlis},\ and\ \citenamefont
  {Sachdev}}]{Patel2023}%
  \BibitemOpen
  \bibfield  {author} {\bibinfo {author} {\bibfnamefont {A.~A.}\ \bibnamefont
  {Patel}}, \bibinfo {author} {\bibfnamefont {H.}~\bibnamefont {Guo}}, \bibinfo
  {author} {\bibfnamefont {I.}~\bibnamefont {Esterlis}},\ and\ \bibinfo
  {author} {\bibfnamefont {S.}~\bibnamefont {Sachdev}},\ }\href
  {https://doi.org/10.1126/science.abq6011} {\bibfield  {journal} {\bibinfo
  {journal} {Science}\ }\textbf {\bibinfo {volume} {381}},\ \bibinfo {pages}
  {790} (\bibinfo {year} {2023})}\BibitemShut {NoStop}%
\bibitem [{\citenamefont {Huang}\ \emph {et~al.}(2019)\citenamefont {Huang},
  \citenamefont {Sheppard}, \citenamefont {Moritz},\ and\ \citenamefont
  {Devereaux}}]{Huang2019}%
  \BibitemOpen
  \bibfield  {author} {\bibinfo {author} {\bibfnamefont {E.~W.}\ \bibnamefont
  {Huang}}, \bibinfo {author} {\bibfnamefont {R.}~\bibnamefont {Sheppard}},
  \bibinfo {author} {\bibfnamefont {B.}~\bibnamefont {Moritz}},\ and\ \bibinfo
  {author} {\bibfnamefont {T.~P.}\ \bibnamefont {Devereaux}},\ }\href
  {https://doi.org/10.1126/science.aau7063} {\bibfield  {journal} {\bibinfo
  {journal} {Science}\ }\textbf {\bibinfo {volume} {366}},\ \bibinfo {pages}
  {987} (\bibinfo {year} {2019})}\BibitemShut {NoStop}%
\bibitem [{\citenamefont {Patel}\ and\ \citenamefont
  {Sachdev}(2019)}]{Patel2019}%
  \BibitemOpen
  \bibfield  {author} {\bibinfo {author} {\bibfnamefont {A.~A.}\ \bibnamefont
  {Patel}}\ and\ \bibinfo {author} {\bibfnamefont {S.}~\bibnamefont
  {Sachdev}},\ }\href {https://doi.org/10.1103/PhysRevLett.123.066601}
  {\bibfield  {journal} {\bibinfo  {journal} {Phys. Rev. Lett.}\ }\textbf
  {\bibinfo {volume} {123}},\ \bibinfo {pages} {066601} (\bibinfo {year}
  {2019})}\BibitemShut {NoStop}%
\bibitem [{\citenamefont {Allocca}(2024)}]{Allocca2024}%
  \BibitemOpen
  \bibfield  {author} {\bibinfo {author} {\bibfnamefont {A.~A.}\ \bibnamefont
  {Allocca}},\ }\href@noop {} {} (\bibinfo {year} {2024}),\ \Eprint
  {https://arxiv.org/abs/2402.18626} {arXiv:2402.18626 [cond-mat.str-el]}
  \BibitemShut {NoStop}%
\bibitem [{\citenamefont {Fournier}\ \emph {et~al.}(2023)\citenamefont
  {Fournier}, \citenamefont {Downey}, \citenamefont {H\'ebert}, \citenamefont
  {Charlebois},\ and\ \citenamefont {Tremblay}}]{Fournier2023}%
  \BibitemOpen
  \bibfield  {author} {\bibinfo {author} {\bibfnamefont {J.}~\bibnamefont
  {Fournier}}, \bibinfo {author} {\bibfnamefont {P.-O.}\ \bibnamefont
  {Downey}}, \bibinfo {author} {\bibfnamefont {C.-D.}\ \bibnamefont
  {H\'ebert}}, \bibinfo {author} {\bibfnamefont {M.}~\bibnamefont
  {Charlebois}},\ and\ \bibinfo {author} {\bibfnamefont {A.-M.}\ \bibnamefont
  {Tremblay}},\ }\href@noop {} {} (\bibinfo {year} {2023}),\ \Eprint
  {https://arxiv.org/abs/2312.08306} {arXiv:2312.08306 [cond-mat.str-el]}
  \BibitemShut {NoStop}%
\end{thebibliography}%

\end{document}